\documentclass[pdflatex,sn-basic]{sn-jnl}

\usepackage{graphicx}%
\usepackage{multirow}%
\usepackage{amsmath,amssymb,amsfonts}%
\usepackage{amsthm}%
\usepackage{mathrsfs}%
\usepackage[title]{appendix}%
\usepackage{xcolor}%
\usepackage{textcomp}%
\usepackage{manyfoot}%
\usepackage{booktabs}%
\usepackage{algorithm}%
\usepackage{algorithmicx}%
\usepackage{algpseudocode}%
\usepackage{listings}%
\usepackage{rotating}           
\usepackage{lscape}				
\usepackage{pdflscape}
\usepackage{threeparttable} 
\usepackage{subcaption}
\usepackage{multirow}
\usepackage{graphicx}  
\usepackage{booktabs}
\usepackage{siunitx}
\usepackage{nicefrac} 
\theoremstyle{thmstyleone}%
\theoremstyle{thmstyletwo}%

\theoremstyle{thmstylethree}%

\begin{document}

\title[Article Title]{Pricing of wrapped Bitcoin and Ethereum on-chain options}


\author{\fnm{Anastasiia} \sur{Zbandut}}



\abstract{This paper measures price differences between Hegic option quotes on Arbitrum and a model-based benchmark built on Black--Scholes model with regime-sensitive volatility estimated via a two-regime MS-AR-(GJR)-GARCH model. Using option-level feasible GLS, we find benchmark prices exceed Hegic quotes on average, especially for call options. The price spread rises with order size, strike, maturity, and estimated volatility, and falls with trading volume. By underlying, wrapped Bitcoin options show larger and more persistent spreads, while Ethereum options are closer to the benchmark. The framework offers a data-driven analysis for monitoring and calibrating on-chain option pricing logic.}




\maketitle

\section{Introduction}\label{SEC:Introduction}

Over the past few years, automated market makers (AMMs) have expanded from spot trading to perpetuals and, most recently, to options. They are core of decentralized finance (DeFi) because they replace the traditional order‐book with a programmatic pricing rule backed by pooled liquidity.
In centralized finance (CeFi), a centralized exchange discovers prices via a limit order book, holds client assets in custody, and relies on off-chain clearing and risk management \citep{cefidefi2021}. In DeFi, these functions are implemented in code where smart contracts hold collateral, compute quotes according to an AMM formula, manage oracles and validators sustain liveness on the blockchain \citep{amm2023,DeFI2022,Rahman2022,eskandari2017}.
Consequently, on-chain option prices are determined algorithmically by four components: (i) the AMM pricing rule and its parameters, (ii) the liquidity pool’s inventory and utilization, (iii) the oracle’s reference index and update frequency, and (iv) protocol fees together with throughput constraints \citep{volkovich2023,priem2022}.

The four components above map directly to the main frictions of on-chain pricing. First, the AMM pricing rule and its parameters include denomination and settlement choices, which change payoff curvature and can introduce basis risk even when the spot market is liquid \citep{alexander2023}. Second, the liquidity pool’s inventory and utilization matter because pooled capital must warehouse option convexity under full collateralization. Moreover, near-instant settlement at block finality reduces netting benefits and can raise funding needs relative to traditional clearing \citep{priem2022}. Third, the oracle’s reference index and update frequency create timing risk. Threshold updates and latency can open short-term spreads to reference prices. Prior feasibility studies and live oracle analyses show that liveness and data delivery are first-order concerns \citep{volkovich2023,eskandari2017}. 
Fourth, protocol fees and throughput constraints (block capacity) limit how quickly the protocol can update quotes or adjust hedges \citep{eskandari2017}. This speed limit matters more for options as their convex payoffs make values very sensitive to volatility. In cryptocurrency markets, the effect is amplified because Bitcoin (BTC) and Ethereum (ETH) returns feature jumps, regime shifts, heavy tails, and persistence. A pricing rule that omits these features can misstate volatility and distort option values \citep{charles2019,bouri2019,wu2021,madan2019,olivares2020}.

Together, these four components describe how the smart contract sets the quote. We compare those quotes with a benchmark price, i.e., the Black--Scholes (BS) price computed with a regime-sensitive volatility estimate. This benchmark is a model-based reference, not the unobservable “true” price. Deviations from it can arise for two reasons. First, residual benchmark model risk, e.g., jump risk a volatility proxy does not fully capture. Second, economically justified AMM compensation for design and operating constraints, e.g., denomination/settlement basis or oracle timing risk. We mitigate the first source by focusing on short maturities and by using regime-sensitive volatility, therefore, the remaining patterns are informative about the second source.
This interpretation has two practical implications. If deviations are systematic along order size, moneyness, maturity, volatility, and liquidity depth (and economically large) they point to concrete calibration targets for the AMM. If deviations are persistent and hedgeable after costs with feasible instruments, they may also indicate potential arbitrage in principle. Our goal, however, is diagnosis and guidance for protocol tuning rather than proposing a trading strategy. In this way, the study informs both the empirical assessment of on-chain option pricing and the design of option-AMM pricing rules.

This paper studies Hegic protocol on Arbitrum chain by using data for wrapped Bitcoin (wBTC) and ETH options from October 24, 2022 to May 21, 2024. Purely on-chain, AMM-based option protocols remain rare and the segment is still small. At the moment of writing, according to \citep{DeFiLlamaOptions}, total value locked (TVL) in DeFi option protocols is about \$100 million where Hegic accounts for \$28.5 million and is the largest options protocol. Hegic is launched on both ETH and Arbitrum blockchains and with respect of it's TVL, about \$1.7 million is on ETH and \$25.3 million on Arbitrum, which indicates that most activity occurs on Arbitrum. This concentration motivates the focus on the Arbitrum chain. Hegic is a long-only, pooled-liquidity protocol that offers American-style call and put options on a discrete strike–maturity grid and prices via a rate-based rule. On the contrary, other on-chain option protocols,  Lyra/Derive and Deri implement BS–style AMM logic driven by oracle inputs and, for Lyra/Derive, an implied volatility (IV) surface. This design difference makes Hegic an informative case for evaluating how an AMM’s pricing rule maps into observed quotes relative to a benchmark BS valuation, and along which variables, i.e., order size, moneyness, maturity, volatility, liquidity depth, the mispricing is most pronounced.

We construct a model-based benchmark valuation by applying the BS model with a regime-sensitive estimate of the underlying volatility, consistent with the nature of cryptocurrency returns (heavy tails, volatility clustering, and regime shifts).
Specifically, we estimate a two–regime MS–AR–(GJR)–GARCH model for the underlying asset. The Markov component accommodates structural breaks and persistence documented for BTC and ETH, while the GJR term captures asymmetric volatility responses within regimes \citep{charles2019,bouri2019,wu2021}.
When the option market offers only a sparse set of quotes across strikes and maturities, i.e., low liquidity depth or intermittent trading,\footnote{Typical for on-chain protocols that list only a few ATM/OTM strikes on discrete maturities.} GARCH-based measures provide a practical proxy for expected volatility and capture large moves in cryptocurrency markets \citep{venter2020}. Forecast studies also show that cryptocurrency volatility is better explained by crypto–specific factors, than by equity–based benchmarks \citep{liang2022}. For these reasons, the BS model with a regime-sensitive volatility serves as a coherent reference price for cross-sectional comparisons \citep{olivares2020,madan2019,CaoCelik2021, Cao2001}. We then define mispricing as the relative difference between the benchmark and the Hegic price and explain it with feasible GLS, accounting for heteroskedasticity, same–day dependence across contracts written on the same underlying. 
The paper reports both statistical significance and economic magnitude.

The main findings are threefold. First, the benchmark price is higher than the Hegic quote on average, except for ETH put options, and the deviation is larger for call options. For wBTC mispricing increases with volatility, while this effect is not statistically distinguishable from zero for ETH. Second, the price deviation increases when the order is larger, the strike is further from the money, the maturity is longer, and the underlying is more volatile. On the contrary, it decreases when underlying trading volume is higher. Third, ETH options are generally closer to the benchmark than wBTC options.
These patterns admit a clear economic interpretation. In a pooled-liquidity AMM, larger orders impose higher marginal inventory and convexity risk on the pool. Far OTM and short-maturity options concentrate that convexity where hedging is most fragile, so quotes tend to move further from the benchmark. On the contrary, deeper markets improve the ability to hedge and bring quotes closer to the benchmark price.
The price deviation varies with market conditions. In high-volatility periods, cross-exchange price differences increase, however, when centralized exchange depth and on-chain transfer activity are higher, these differences decrease \citep{kristoufek2023,volkovich2023}.
Common long-run factors and time-varying error correction imply that wBTC and ETH deviations can co-move in some periods \citep{keilbar2021,leung2019}.
During large market moves or liquidity squeezes, prices tend to adjust first in the futures market while spot follows. The delayed price discovery or contract design can cause temporary differences between futures and spot prices during big market moves \citep{akyildirim2020}.
Additionally, market-wide efficiency in major cryptocurrencies has improved over time and is strongest when liquidity and money flows are high \citep{lopez2021,letran2020,fernandes2022,mokni2025,aslam2023}. The results of the paper point to protocol design and market plumbing as the primary drivers of systematic deviations from the benchmark.

This paper complements the study of \citet{AndolfattoNaikSchoenleber2024}. Their focus is the implied–volatility (IV) difference between Lyra (on-chain) and Deribit (off-chain) options. They document the an on–chain IV premium is related to retail net buying of call options and to trading–volume shocks. The authors also examine the performance of a volatility–premium trading strategy. The focus of this paper is different, it analyses the on-chain pricing rule at the transaction level. 
This trade-by-trade comparison isolates how the AMM prices options and where the price deviates from a model-based reference. Therefore, the paper investigates whether the deviation co–moves with the specific rules that an option AMM follows.
The two paper are, therefore, complementary. Cross–exchange IV comparisons show how on-chain markets as a whole differ from centralized order books. 
Protocol-specific price diagnostics show how the AMM’s design maps market conditions into prices.


The contribution of this paper is both empirical and practical. First, with respect to volatility modelling/forecasting, we show that a two–regime MS–AR–(GJR)–GARCH model delivers a credible short-term volatility estimate for on-chain option valuation and we document asset-specific regime features (BTC vs.\ ETH) that are informative for forecasting and risk control. Second, with respect to protocol design, we map the documented price deviations into actionable calibration levers for Hegic-type AMMs. Third, with respect to the literature on on-chain option AMMs, we provide the first transaction-level evaluation of a rate-based option AMM and, thereby, advancing the empirical basis for AMM design in on-chain options.

The rest of the paper is organized as follows: Section \ref{SEC:litreview} reviews related work on cryptocurrency market efficiency, volatility regimes and breaks, and DeFi microstructure. Section \ref{SEC:EmpAnalysis} outlines the methodology design in Subsection \ref{SubSEC:Methodology}, presents the design of the Hegic protocol and data in Subsection \ref{SubSEC:Data}. Section \ref{SubSEC:EmpResults} reports the results and offers discussion. Section \ref{sec:conclusion} concludes the paper with implications and limitations.

\section{Related literature}\label{SEC:litreview}

We group the literature into four strands that motivate the methodology choices of this paper: (\ref{subsec:market}) market efficiency; (\ref{subsec:vola}) volatility regimes, jumps, and structural breaks; (\ref{subsec:options}) option microstructure, denomination, and hedging; and (\ref{subsec:defi}) DeFi plumbing, i.e., AMMs, oracles, bridges, and protocol design. 

\subsection{Market efficiency}\label{subsec:market}

\citet{letran2020} examine weak–form efficiency for BTC, ETH, XRP, LTC, and EOS during 2013–2019 using the Adjusted Market Inefficiency Magnitude and find that efficiency improves during 2017–2018, with event–driven setbacks. The authors find that LTC is most and XRP least efficient on average. \citet{lopez2021} apply linear and nonlinear tests to BTC, LTC, ETH, XRP, XLM, and XMR and report that BTC, LTC and ETH become more efficient in later subsamples, consistent with the Adaptive Market Hypothesis. 
Earlier work also documents time variation in efficiency and the return–risk trade-off. \citet{Bariviera2017} compute rolling Hurst exponents for BTC returns and volatility during 2011–2017 and show that return efficiency improves over time while volatility retains long memory and clustering. Focusing on the 2013 crash, \citet{Bouri2017} estimate asymmetric GARCH models on daily BTC data and document an inverse leverage effect, i.e., volatility reacts more to positive than to negative return shocks, with dynamics that differ across the pre- and post-crash windows.
Using entropy–information methods for BTC, ETH, ADA, BNB, and XRP prices during 2018–2021, \citet{fernandes2022} find high and stable efficiency before and during COVID–19, with ETH exhibiting the least variation.
Multifractal evidence further clarifies the state dependence of efficiency. \citet{aslam2023} apply multifractal detrended fluctuation analysis to daily prices of ADA, BNB, BTC, ETH, LTC, and XRP up to early 2023 and rank assets by the width of the multifractal spectrum and a long-memory index. They find a lower weak-form efficiency, for BTC and LTC when multifractality is high. On the contrary, a stronger herding during crisis windows for ADA and BNB when multifractality is low. Therefre, they argue that efficiency varies with market conditions and across assets. Complementing this finding, \citet{mokni2025} measure efficiency using the Adjusted Market Inefficiency Magnitude (AMIM) for 2016–2023 and explain it with quintile regressions on global factors, e.g., financial stress, equity benchmarks, substitutes, e.g., commodities, and internal crypto variables, e.g., liquidity, volatility, money flows. They report that greater liquidity and stronger money flows are associated with lower AMIM, i.e., higher efficiency, whereas financial stress raises AMIM with effects that vary across quintiles. These studies imply that efficiency is time-varying and improves with depth and activity. Therefore, this paper treat persistent on-chain price deviations as design- and state-driven and investigates whether they decrease when underlying volume and market liquidity increase.

\subsection{Volatility regimes, jumps, and structural breaks}\label{subsec:vola}

A large strand of literature shows that cryptocurrency volatility is clustered, asymmetric, and state dependent, with jumps and structural breaks. Using daily date for BTC prices over the early market years up to 2016, \citet{Katsiampa2017} compare GARCH variants and show that GARCH–type specifications fit volatility better than simple historical measures. For BTC risk measurement, \citet{Stavroyiannis2018} estimate VaR and related metrics on daily data and document heavy tails, underscoring the relevance of fat–tailed errors for the option risk estimation. Focusing on regime changes, \citet{Thies2018} apply Bayesian change–point methods to BTC daily returns during 2012–2018 and identify multiple volatility regimes. Over an earlier sample,, during 2011–2013, \citet{Ardia2019} compare Markov–switching AR–GARCH models and find that a two–state MS–AR–GJR–GARCH with skewed–$t$ innovations performs best, confirming both regime switching and asymmetric responses. 
The literature on the cross–asset evidence is consistent with this view. Using daily data for BTC, DASH, LTC, and XRP over 2014–2018, \citet{charles2019} first detect and filter jumps and control for variance breaks. Once these are accounted for, GARCH inference changes significantly, IGARCH($t$) best fits BTC, DASH, and LTC, FIGARCH($t$) fits XRP, and leverage effects disappear. For BTC over 2010–2016, \citet{bouri2019} document highly persistent shocks at the full–sample level, mean reversion within break–defined subsamples, and long memory in volatility, highlighting the role of structural breaks for persistence measurement. 

A part of stand argues that the window length matters as well. Across BTC, ETH, and XRP over 2011–2020, \citet{wu2021} show that out–of–sample VaR from MS–GARCH is sensitive to the estimation window even when within–regime parameters are stable. They find BTC performing best around a 400–day window and ETH closer to 600 days. When listed option quotes are rare, \citet{venter2020} construct GARCH–generated 30/60/90–day volatility indices for BTC and CRIX and validate them by pricing BTC options. The authors find that the indices track jumps and can invert at large shocks, supporting GARCH–type estimates as forward–looking proxies. The literature on forecasting points to crypto–specific drivers. Using the data over 2013–2019, \citet{liang2022} find that commodity volatility and investor attention (Google Trends) improve BTC volatility forecasts, while VIX adds little. Complementing these results, \citet{chenhardle2020} estimate stochastic–volatility–with–jumps models on high–frequency intraday and daily BTC and show pervasive return and variance jumps and co–jumps that steepen short–maturity smiles. They find that diffusion–only benchmarks understate tail risk.
These studies show that ignoring jumps and breaks biases volatility estimates, and  stable regime identification benefits from long samples that include multiple market conditions (booms, busts, stress events). Accordingly, we estimate a two–state MS–AR–(GJR)–GARCH on a long daily window for each underlying. This choice follows the window–sensitivity evidence in \citet{wu2021}, the option–pricing validation of GARCH indices in \citet{venter2020}, and the crypto–specific forecasting drivers in \citet{liang2022}.

\subsection{Option microstructure, denomination, and hedging}\label{subsec:options}

Empirically studies argue that simple diffusion models, e.g., the BS model with constant volatility, do not fit observed BTC option surfaces. Particularly, the pronounced smiles and short–maturity curvature. 
Using ten BTC option surfaces\footnote{The BTC option surface is built on an index aggregating quotes from six BTC–USD spot exchanges, i.e., Bitfinex, Bitstamp, GDAX (Coinbase Pro), Gemini, Itbit, and Kraken. The options are European, cash–settled in BTC, with USD as the numéraire.} collected weekly from June 29 to August 31, 2018,  \citet{madan2019} show that stochastic–volatility and time–changed L\'evy models outperform diffusion-only benchmarks. They report that the conic-finance implied liquidity metric rises with maturity and falls as options move further out of the money. Thus, liquidity is thinnest for short-maturity, far OTM options.
Complementing this funding, \citet{olivares2020} calibrate a mean-reverting jump–diffusion under an Esscher-selected risk-neutral measure on daily BTC–USD data from January 2011 to June 2018 and reject diffusion-only specifications. The author finds that return and variance tails require jumps, which is most consequential for short maturities.
\citet{alexander2023} develop pricing for direct, inverse, and quanto–inverse cryptocurrency options and show that denomination and index–based settlement change payoff curvature and greeks. These design choices create a systematic deviation between what the option pays and what can be replicated with traded hedges, indicating that replication is only approximate. \cite{CaoCelik2021} do not specify the options' nature, however contributes to the field by deriving the analytical formula for pricing BTC options. It is based on the equilibrium model of \cite{Cao2001} where money supply and aggregated dividend follow jump-diffusion processes. The authors assume BTC as foreign currency in a small open economy and empirically find that the BS model underprices BTC options compared to their model. Although the authors do not differentiate between off-chain and on-chain options, they find that IV is sensitive to jumps in money supply and suggest that additional pricing risk should be considered when valuing cryptocurrency options.

Using one–minute data around the launch of U.S. BTC futures (December 2017–February 2018), \citet{akyildirim2020} show that price discovery often originates in the futures market and that structural breaks coincide with major news and regulatory events. Short–term spreads between futures and spot may be reflected in oracle-reported prices and, in turn, into on–chain quotes. \citet{sebastiano2020} uses daily data and find that hedging BTC with BTC futures delivers large reductions in variance and semivariance, whereas cross–hedging ETH,LTC, or XRP with BTC futures is weak and can even increase expected shortfall. 
Using a large cross–section of major cryptocurrencies with daily data up to 2019, \citet{keilbar2021} report four long–run cointegrating relations and introduce a COINtensity–VECM in which the strength of error correction varies over time. The error–correction loadings are the largest during the 2017–2018 bubble, so deviations from the long–run cointegrating relations close more quickly.
\citet{leung2019} construct cointegrated spreads, using Coinbase daily prices for BTC, ETH, LTC, and BCH from December 20, 2017 to June 20, 2018, and show that they are tradable in–sample without costs, However, the performance of this strategy is sensitive to realistic transaction frictions. From the bear-market evidence, \citet{kyriazis2019} show that most major cryptocurrencies are complementary to BTC/ETH/XRP and that principal coins offer no hedging benefits in distressed times. The authors argue that information-criterion selection favors asymmetric and nonlinear GARCH variants for many assets, indicating time-varying co-movement. This findings explains why on-chain AMM deviations can become systemic during stress.

This literature strand implies larger price deviations for bigger orders and for options further out of the money or nearer to maturity. On the contrary, it suggests decreasing price deviations with greater liquidity depth, market activity, and the possibility of cross-asset co-movement of deviations in particular market states.

\subsection{DeFi plumbing: AMMs, oracles, and protocol design}\label{subsec:defi}

On-chain pricing is constrained by how protocols encode trading, data, and settlement. In an early feasibility study on ETH, \citet{eskandari2017} implement a fully collateralized option and two oracle designs: (i) callback oracles that trigger contract updates on events, and (ii) per–block oracle that write prices each block. Their experiments show that while the option logic is straightforward to encode, oracle delivery, gas costs, and liveness are binding: callbacks can miss or delay updates, whereas per–block posting improves liveness at the cost of trusting an oracle. \citet{volkovich2023} compare Mycelium, a centralized exchange oracle on Arbitrum, with Chainlink, a cross-chain decentralized oracle network, around two high–volatility windows in around FTX collapse (November 2022). 
They measure the percentage deviation between each oracle and a three-exchange median (Binance, FTX, Bitfinex). In calm periods, the Mycelium oracle updates quickly  with average latency $\approx 0.3$\,seconds and stays close to the median and, on high-volatility days, the deviation widens. On the contrary, Chainlink’s 2.5\% deviation threshold can delay updates. Consequently, those deviations across centralized exchanges can incfluence the oracle feed and, in turn, into on-chain quotes.

\citet{Rahman2022} survey option protocols such as Lyra, Deri, Vega, and Thales (primarily on ETH/L2s) and document the key design choices that differentiate them from centralized venues are: AMM pricing rules, oracle providers, liquidity architecture e.g., peer–to–pool vs.\ peer–to–peer, and governance/incentive tokens. These choices impact inventory, funding, and data–quality risks that affect quoted premiums. 
\citet{priem2022} examine the OTC derivatives life cycle on distributed ledgers and show that near-instant settlement reduces multilateral netting, thereby increasing gross funding needs for long-term positions. They also argue that smart contracts do not, by themselves, resolve issues of legal enforceability, privacy, or cross-platform interoperability. These frictions provide a reason why on-chain quotes can persistently differ from the benchmark.
\citet{McCorry2021} classify bridge validation designs and highlight core trade-offs among security assumptions, latency, and interoperability. Bridge choices determine how quickly and safely assets and price information can move across domains. They interact with oracle update cadence, influencing the timing of settlement and hedging on-chain.
Reviewing empirical work on stablecoins, \citet{Ante2023} summarize evidence on adoption, liquidity, and peg stability. The authors emphasize that market distress can lead to temporary depegs and liquidity strains. When option protocols settle in stablecoins, these dynamics can influence realized payoffs, liquidity provision, and risk management.

The above mentioned evidence implies that differences between an AMM quote and the benchmark price can arise from protocol setting. Accordingly, we include explanatory variables that map directly to those mechanisms: order size (inventory and funding pressure in a pooled pool), moneyness and maturity (option convexity and hedgeability), underlying volatility (data/throughput stress and hedging difficulty), and liquidity depth (external hedging capacity). 

The combined findings of the literature strands indicate that since major cryptocurrency markets are often informationally efficient for long periods, persistent on–chain deviations should be traced to protocol setting and market conditions rather than to a failure of price discovery. Because volatility is jumpy and regime dependent, the volatility estimation should cover heavy tails and short–term convexity. Since denomination/settlement, hedging capacity, and oracle design affect replication and inventory risk, deviations should be larger when orders are bigger, strikes are further from the money, maturities are short, or volatility is elevated, and should shrink when underlying volume and on–chain activity increase in the market. These predictions map directly into our empirical specification and the interpretation of economic magnitudes.

\section{Empirical analysis}
\label{SEC:EmpAnalysis}
This section outlines how we construct a regime–sensitive volatility input and a benchmark price to evaluate Hegic quotes, and how we then measure and explain mispricing.

\subsection{Methodology}\label{SubSEC:Methodology}

The methodology starts with constructing a regime–switching volatility input using  BTC and ETH time series and only then prices options with the \citet{BlackScholes1973} model. The resulting benchmark prices are compared to Hegic quotes at the transaction timestamp, and the cross–section of price deviations is explained with a feasible GLS (FGLS) regression. Each step is motivated by the empirical finding on of cryptocurrency returns properties and by the protocol’s microstructure.

\paragraph{Step 1: Regime identification}
We model daily rates of return \(r_t\) with a two–regime Markov–Switching autoregression (MS-AR) as:
\begin{align}
r_t=\mu_{\rho_t}+\sum_{i=1}^{a}\phi_{\rho_t,i}\,r_{t-i}+\varepsilon_{\rho_t,t}, 
\qquad \varepsilon_{\rho_t,t}\sim \mathcal{N}\!\bigl(0,\sigma^2_{\rho_t}\bigr)
\end{align}
where \(\rho_t\in\{0,1\}\) denotes the regime at time \( t \) and follows a first-order Markov chain with transition probabilities
\(p_{ss'}=\Pr(\rho_t=s'\mid \rho_{t-1}=s)\) for \(s,s'\in\{0,1\}\).
We estimate the model by maximum likelihood and get, for each day \(t\),
smoothed state probabilities as:
\begin{align}
\tilde p_t(s)=\Pr(\rho_t=s\mid \mathcal{F}_T),\qquad s\in\{0,1\}
\end{align}
where \(\mathcal{F}_T\) is all information up to the end of the sample. We then sort days into regimes where we assign day \(t\) to the state with the higher smoothed probability as:
\begin{align}
\hat{\rho}_t=\arg\max_{s\in\{0,1\}} \tilde p_t(s)
\end{align}
If \(\max_s \tilde p_t(s)\le 0.5\), we treat \(t\) as uncertain, this avoids noisy classifications. The MS-AR is label-invariant, so after estimation we name the regimes by their level of variances. The regime with the larger variance is labeled high-volatility, the other low-volatility.
We report the transition matrix, expected durations, and the smoothed probability plots which provides the probability of being in the certain regime and how persistent the regimes are.

\paragraph{Step 2: Regime–conditional volatility estimation}
Within each regime identified by the MS--AR model, we fit a regime–conditional GJR--GARCH process with skewed--$t$ distribution to capture volatility clustering and asymmetry \citep{GJR1993,Ardia2019}:

\begin{align}\label{eq:GJR-GARCH}
  \epsilon_{\rho_t,t} &\sim \text{skewed-}t\!\left(0,\ \sigma^2_{\rho_t,t},\,\eta,\lambda\right),\\
  \sigma^2_{\rho_t,t} &= \omega_{\rho_t} \;+\; \sum_{i=1}^{p}\alpha_{\rho_t,i}\,\cdot\epsilon^2_{\rho_t,t-i}
      \;+\; \sum_{i=1}^{o}\gamma_{\rho_t,i}\,\cdot\epsilon^2_{\rho_t,t-i}\,\cdot I(\epsilon_{\rho_t,t-i}<0)
      \;+\; \sum_{j=1}^{q}\beta_{\rho_t,j}\,\cdot \sigma^2_{\rho_t,t-j}
\end{align}
For each cryptocurrency and regime we select $(p,o,q)$ by the Bayesian Information Criterion (BIC) over $p,q\in\{1,\dots,10\}$ and $o\in\{0,\dots,5\}$. If $o=0$, the specification reduces to standard GARCH. We report AIC/BIC, Ljung--Box and Engle’s ARCH statistics on standardized residuals, as well as the estimated degrees of freedom and skewness, to document fit and remaining dynamics. The conditional variance path $\sigma^2_{\rho_t,t}$ provides a time–varying volatility input aligned to each options' purchase date. We convert the daily conditional variance to annualized volatility.

\paragraph{Step 3: Benchmark price estimation}
Given $\sigma_t$ from step 2, we compute the BS benchmark price for call $C$ and put $P$ options as:
\begin{align}\label{eq:bs}
  C &= S \cdot N(d_1) - K \cdot e^{-r_f\cdot T} \cdot N(d_2) \\
		P &= K \cdot e^{-r_f\cdot T} \cdot N(-d_2) - S \cdot N(-d_1)
\end{align}
where $S$ denotes the price of the underlying, $K$ denotes the strike price, $r_f$ denotes the risk free rate, $T$ denotes maturity, $d_1 = \frac{\ln\left(\frac{S}{K}\right) + \left(r + \frac{\sigma^2}{2}\right)\cdot T}{\sigma\cdot \sqrt{T}}$ and $d_2 = d_1 - \sigma \cdot \sqrt{T}$. We treat BS with a regime–switching volatility as a reference price suitable for short maturities, not as a full structural model of the risk–neutral measure \citep{madan2019,olivares2020,venter2020,liang2022}. This design matches the single–purchase nature of Hegic contracts.

\paragraph{Step 4: Mispricing measurement}
For each option $j$, we define the relative price deviation, i.e., mispricing, as:
\begin{align}
  \Delta\text{price}_{j} \;=\; \frac{C(P)_{j}-O_{j}}{O_{j}}
\end{align}\label{eq:misprice}
where $O_j$ is the Hegic quote and $C(P)_j$ is the benchmark price for call and put, respectively. A positive value indicates that the benchmark exceeds the Hegic quote, whereas a negative value indicates the reverse. All inputs are aligned at the exact trade timestamp. 

\paragraph{Step 5: Cross-sectional analysis}

We explain cross-sectional variation in $\Delta\text{price}_j$ with trade-time variables that mirror the AMM pricing rule and market condition. Order size (\emph{Amount}) captures inventory and funding pressure in a pooled AMM. Larger trades generate convexity risk in the pool and are costlier to absorb, especially with near-instant settlement \citep{priem2022,Rahman2022}. Moneyness (\emph{Strike}) captures how far the option is from the money. Far OTM options are harder to replicate and trade in illiquid markets. 
Maturity (\emph{Maturity}) matters for two opposing reasons. Options near maturity exhibit high gamma, which makes hedging more complex, whereas options far from maturity lock collateral for longer and increase funding needs \citep{priem2022}. 
The underlying rate of return (\emph{Return}) and trading volume (\emph{Volume}) are proxies for liquidity depth and trading activity. The greater the liquidity depth and on-chain activity the smaller are cross-exchange price deviations \citep{kristoufek2023}. We also include the underlying volatility $\sigma_t$, since it tracks jumps and provide forward-uncertainty proxies \citep{venter2020}. Finally, dummies for option kind (call=1) and type (ATM=1) allow for side-specific pressure, e.g., retail net call buying on-chain \citep{AndolfattoNaikSchoenleber2024}. 

Since many options are bought on the same day on the same underlying, the regression errors can be heteroskedastic (their variance changes with trade characteristics such as order size or moneyness) and within–day correlated (trades share the same market conditions). OLS coefficients remain unbiased but are inefficient, and their standard errors can be misleading. We therefore use two–step feasible GLS (FGLS) \citep[pp.~204--210]{CampbellLoMacKinlay1998} to model the error variance, reweight observations, and report standard errors that remain valid even when residual variances differ across trades. The first step is a baseline fit where we estimate the cross–sectional model with standard OLS as:
\begin{align}
\Delta\text{price}_j
  &= \beta_0
   + \beta_{\text{amount}}\cdot\text{Amount}_j
   + \beta_{\text{strike}}\cdot\text{Strike}_j
   + \beta_{\text{maturity}}\cdot\text{Maturity}_j\nonumber\\
  &\quad
   + \beta_{\text{return}}\cdot\text{Return}_{t,j} 
   + \beta_{\text{volume}}\cdot\text{Volume}_{t,j}
   + \beta_{\text{vol}}\cdot\text{Volatility}_{t,j}\nonumber\\
  &\quad
   + \beta_{\text{kind}}\cdot\text{Kind}_j
   + \beta_{\text{type}}\cdot\text{Type}_j
   + \xi_j \label{eq:ols_text}
\end{align}
From this step, we collect residuals \(\hat\xi_j\) which contain the behavior of how error variance varies across trades. The next step in the cross-sectional analysis is to estimate the variance model and weights. For this, we square the collected residuals and explain their dispersion with the same regressors via auxiliary regression as:
\begin{align}
\hat\xi_j^{\,2}
  &= \gamma_0
   + \gamma_{\text{amount}}\cdot\text{Amount}_j
   + \gamma_{\text{strike}}\cdot\text{Strike}_j
   + \gamma_{\text{maturity}}\cdot\text{Maturity}_j\nonumber\\
  &\quad
   + \gamma_{\text{return}}\cdot\text{Return}_{t,j} 
   + \gamma_{\text{volume}}\cdot\text{Volume}_{t,j}
   + \gamma_{\text{vol}}\cdot\text{Volatility}_{t,j}\nonumber\\
  &\quad
   + \gamma_{\text{kind}}\cdot\text{Kind}_j
   + \gamma_{\text{type}}\cdot\text{Type}_j
   + \nu_j \label{eq:aux_text}
\end{align}
The fitted values form the diagonal elements of the omega matrix \(\Omega\). This variance-covariance matrix, which incorporates the heteroskedasticity pattern. This variance-covariance matrix incorporates the heteroskedasticity pattern of the data. Observations predicted to have large residual variance, e.g., big orders or far–OTM strikes, receive less weight in FGLS regression (the final step); low–variance observations receive more weight. This directly addresses heteroskedasticity and improves estimates efficiency.
The final step is the GLS regression, estimated as:
\begin{align}
\hat\beta_{\text{GLS}}
  = (X^\top\cdot \Omega^{-1} \cdot X)^{-1} \cdot X^\top\cdot  \Omega^{-1}\,\cdot \Delta\text{price} \label{eq:gls_text}
\end{align}
where \(X\) contains the same regressors as in equation \eqref{eq:ols_text}. This step is equivalent to weighted least squares where each trade is weighted by the inverse of its predicted error variance. When error variance differs across trades, these GLS estimates use the information more efficiently than OLS and typically yield more reliable standard errors. We report heteroskedasticity– and autocorrelation–consistent (HAC) standard errors \citep{Newey1987}, which are robust to unequal variances and short–term correlation. We also report Wald tests for joint significance, the adjusted \(R^2\), the condition number, and variance–inflation factors (VIFs) to document multicollinearity.

\subsection{Data}
\label{SubSEC:Data}

Hegic is a peer-to-pool option AMM deployed on ETH and Arbitrum blockchains. It lists American-style options (called hedge contracts) on wrapped wBTC and ETH, and allows only long positions. The options' counterparty is a pooled liquidity vault funded in stablecoins (DAI, USDC, USDT).Premiums paid by buyers accrue to the pool, and losses are allocated to liquidity providers in proportion to each provider’s share of the pool. The protocol relies on an external price oracle (Chainlink) for the underlying, settles in DAI, and routes swaps through a decentralized exchange (Uniswap). When adding liquidity to the poool, liquidity providers receive ERC-20 token, e.g., writeDAI, and a portion of the pool is locked until option expiry. Withdrawals are otherwise subject to available unlocked liquidity and a queueing mechanism. A portion of idle DAI can be wrapped into CHAI to earn interest via MakerDAO, i.e., additional incentive for the liquidity provision \citep{Hegic2020}.

The AMM pricing rule on Hegic is unique and rate-based. For a given strike $K$ and maturity $T$ from the protocol’s discrete grid, the quoted option premium is calculated as:
\begin{align}
    O = v(K,T)\cdot S
\end{align}
where $v(\cdot)$ is a predefined rate. Buyers pay an additional settlement fee is calculated as:
\begin{align}
    s = \text{Amount}\times O \times r_s
\end{align}
where ATM options carry a $1\%$ settlement rate and OTM options $0.5\%$. The rate of strike grid is discrete and symmetric around the spot. For calls, available strikes are at $100\%,110\%,120\%,130\%$ of the oracle price and, for puts, at $100\%,90\%,80\%,70\%$ with maturities spanning from 7 to 90 days. Hegic does not offer in-the-money options. 

This design contrasts with other on-chain option protocols. Lyra uses a BS–style surface from an external oracle (Block Scholes) and splits risk across a collateral pool and a delta-hedging pool. The protocol offers maturity in four fixed expires, i.e., 7/14/21/28 days. \citep{Lyra2023,Rahman2022} Deri offers everlasting options (BS model with the $T\!\to\!\infty$) where positions are balanced through continuous funding payments and a proactive market–making algorithm that adjusts to oracle quotes. \citep{Deri2023,EverlastingOptions2021}. Additionally, the protocol is deployed across multiple chains which broadens access and can lower user costs on L2s. However, it can also fragment liquidity across exchanges and introduce bridge-related security risk. Moreover, oracle parameters are chain specific, so update timing can also differ. The net effect on depth and price quality is, therefore, ambiguous and state dependent \citep{Rahman2022,McCorry2021,priem2022,volkovich2023}. 

Unlike BS–style AMMs, Hegic calculates quotes from a static rate table on a discrete strike–maturity grid, where users restricted to buying options, i.e., only long positions. 
The level of the quote depends primarily on the rate schedule and the execution depends on available pool liquidity.
In Hegic, the liquidity pool is the sole writer of all options, and both premia and payouts are shared across liquidity providers in proportion to their pool share. When a liquidity provider also buys an option from the same pool, part of the premium they pay flows back to them through their liquidity provider entitlement, and part of any eventual payout is borne by them for the same reason. Economically, this is equivalent to buying a smaller position from an external counterparty. The liquidity provider’s effective premium paid and effective payoff are both reduced by the fraction of the pool they own. Two practical implications follow. First, liquidity provider purchases partially offset the pool’s natural short-convexity exposure without withdrawing liquidity, because the buyer and the holder of the option are partly the same economic agent. Second, a liquidity provider who buys an option from the pool does not create a free gain. Because the pool is the counterparty, part of the premium the liquidity provider pays is redistributed back to them through their pool share, but the same share also funds a proportion of any future payout if the option finishes in the money. After protocol fees and the fact that premia are shared with all liquidity providers, the pay back is incomplete. Economically, such a trade simply rebalances the liquidity provider’s net short-gamma exposure, i.e., reducing risk taken via the pool rather than generating arbitrage profits.

If the Hegic rate \(v\) is not adjusted when market volatility or order pressure changes, the quoted premium \(O\) can drift from a benchmark price \(P^{\text{BS}}\). When \(v\) is set too low for current risk, quotes fall below the benchmark and the pool is under-compensated for the convexity it sells (options are “too cheap” for buyers). When \(v\) is set too high, quotes exceed the benchmark and order flow usually declines, but filled trades pay higher premia to the pool. In practice, if \(O<P^{\text{BS}}\) a trader can buy on Hegic and hedge delta in spot or futures off-protocol; if \(O>P^{\text{BS}}\), the short position must be taken on another exchange because Hegic does not allow users to write options. Execution costs, hedge basis, oracle timing, and available depth limit any cross-exchange strategy and help explain why price deviations can persist \citep{alexander2023,priem2022,volkovich2023}. These motivate our explanatory variables and point to calibration possibilities for a rate-based AMM.

Fully on–chain option AMMs are still a small segment of DeFi. At the time of writing, TVL in DeFi option protocols is about \$100\,million where Hegic accounts for \$28.5\,million and is the largest option AMM by TVL. Within Hegic, \$25.3\,million is locked on Arbitrum and \$1.7\,million on ETH, so most activity occurs on Arbitrum. Therefore, we analyze all Hegic option trades on Arbitrum between October 24, 2022 and May 21, 2024, which is the full history available on that chain at the time of writing.

\begin{sidewaysfigure}[htbp]
  \centering
  \caption{Sankey chart — wBTC option holders}
  \label{FIG:BTC_sankey}
  \includegraphics[width=\textwidth]{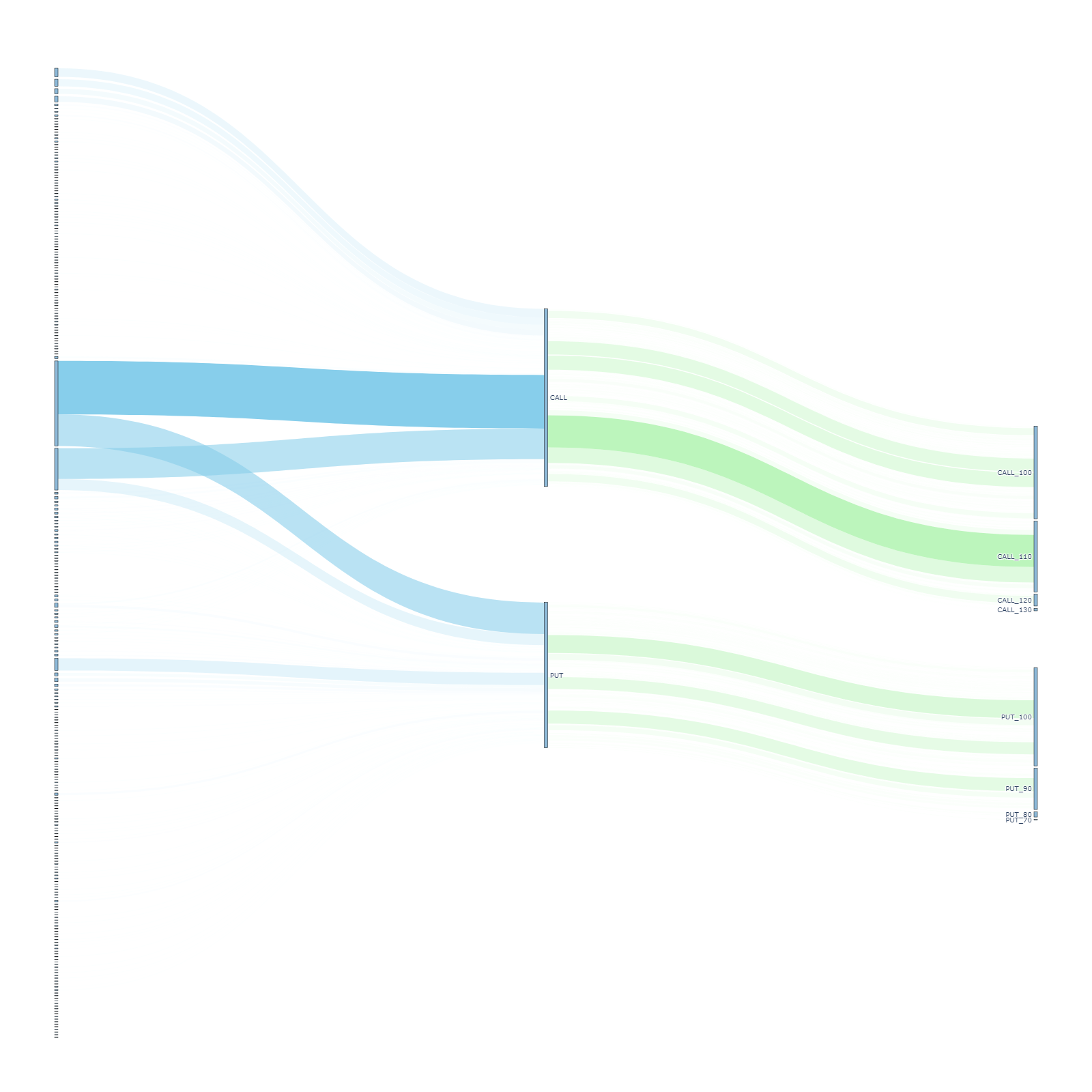}
\end{sidewaysfigure}

\begin{sidewaysfigure}[htbp]
  \centering
  \caption{Sankey chart — ETH option holders}
  \label{FIG:ETH_sankey}
  \includegraphics[width=\textwidth]{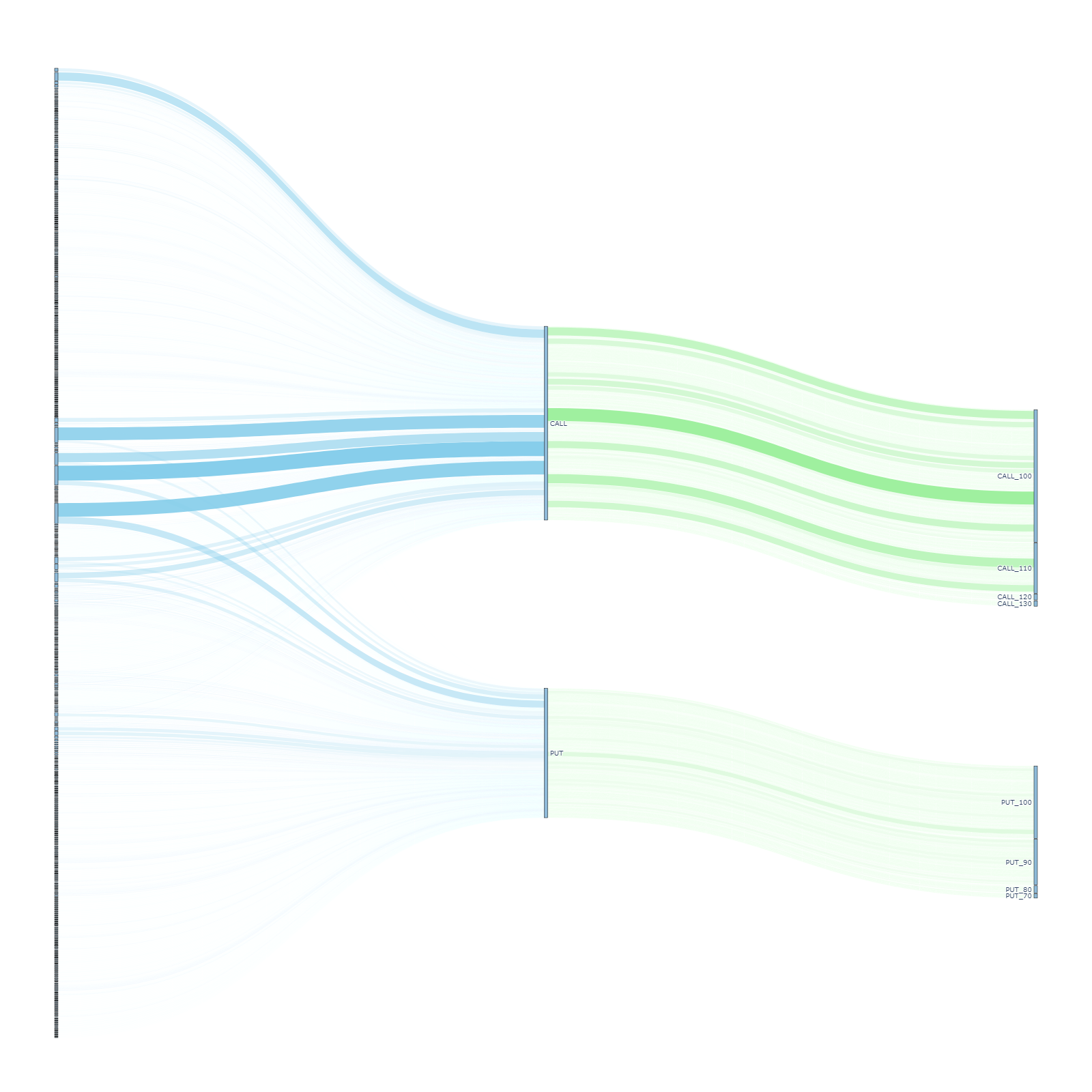}
\end{sidewaysfigure}

Unlike centralized exchanges, decentralized exchange trades are recorded on–chain with public excess. Each observation includes the option type, strike, maturity, amount (fractional units allowed), the paid premium (in stablecoin), and the exact block timestamp, as well as the buyer address and transaction hash. 
Figures \ref{FIG:BTC_sankey}–\ref{FIG:ETH_sankey} show Sankey charts of purchased options by unique address and summarize protocol participation. For wBTC, 1{,}315 options were bought by 275 addresses, with roughly 40\% of all contracts purchased by two addresses.\footnote{0x7...d3d address bought 346 options and 0x7...74c address bought 150 options.} The split is 723 call options (377 ATM options) and 592 put options (400 ATM options). For ETH, 2{,}775 options were bought by 728 addresses, with about 40\% purchased by 18 addresses. ETH order flow is more dispersed than wBTC. ETH totals are 1{,}664 calls (1{,}137 ATM) and 1{,}111 puts (621 ATM). The prevalence of ATM options is consistent with leveraged directional views which align with on–chain evidence in \citet{AndolfattoNaikSchoenleber2024}.

Table~\ref{tab:DescriptiveStats} reports summary statistics by the underlying and kind. Fractional order sizes exhibit very small minimum and mean premiums are of similar level across buckets, with higher dispersion for wBTC reflecting its higher price level. By design, strikes reflect a discrete grid (ATM and fixed OTM steps), and maturities cluster at 7–90 days with similar medians across buckets. ETH options show larger Amount quartiles than wBTC, consistent with deeper trading activity shown in the figure \ref{FIG:ETH_sankey}. Across both underlying, call options show higher mean volumes than puts, in line with speculative demand found in \citet{AndolfattoNaikSchoenleber2024}.

\section{Results}
\label{SubSEC:EmpResults}

We begin by constructing the volatility input for wBTC and ETH options.  For this, the time-series history of daily prices for BTC and ETH prices from December 7, 2018, until May 25, 2024, are retrieved from CryptoCompare via API. This time horizon includes distinct market conditions that are important for identifying regimes, i.e., the COVID-19 shock in March~2020, the 2020–2021 bull run associated with “DeFi Summer,” and major policy episodes, e.g., EU MiCA negotiations. A long history that covers both calm and stressed periods is necessary for reliable regime detection. Otherwise, regime probabilities and state-conditional dynamics are weakly identified. Prior research documents structural breaks, long memory, and regime changes in cryptocurrency volatility, and shows that inference and tail-risk forecasts are sensitive to the estimation window \citep{Thies2018,Ardia2019,charles2019,bouri2019,wu2021}. Consistent with this evidence, we use roughly \(\sim\)2{,}000 daily observations to estimate a two-regime MS–AR–(GJR)–GARCH: the Markov-switching autoregression assigns each day to a low- or high-volatility regime, and, within each regime, a GJR–GARCH with skewed-\(t\) innovations captures volatility clustering and asymmetry. This provides the rationale for the regime–sensitive volatility input and a longer time horizon than the option sample.

Table \ref{TAB:summaryunderlying} summarizes the statistics of BTC and ETH rates of return. Mean rates of return for both samples are positive, with around four percentage points higher for ETH. This is likely due to the continuously growing DeFi ecosystem, where ETH is the preferable blockchain and plays a vital role in various on-chain applications. ETH rates of return also exhibit higher standard deviation and minimum and maximum values, suggesting significant price variations. This, in turn, again highlights the dominant role of ETH in the ecosystem, where an increased amount of projects deployed on its chain and a high transaction flow lead to more frequent price changes. Both BTC and ETH time series are left-skewed, suggesting events with a more pronounced price decline. Similarly, high kurtosis values indicate fat tails, which emphasize extreme values in rates of return. 

Table \ref{TAB:markov} reports estimates from two–regime MS–AR models in which the intercept, autoregressive coefficients, and innovation variance may differ by regime.  Robust standard errors (HAC) are in parentheses and statistical significance is based on $t$–statistics: $^{***}$, $^{**}$, and $^{*}$ denote the 1\%, 5\%, and 10\% levels, respectively. Parameters are obtained by maximum likelihood using the limited-memory BFGS (L-BFGS) algorithm, which approximates the inverse Hessian without storing the full matrix. The table also reports heteroskedasticity-robust standard errors in parentheses \citep{White}. The autoregressive order is eight for BTC and ten for ETH (selected from PACF figures \ref{FIG:pacf_btc} and \ref{FIG:pacf_eth}). Regime labels are assigned by variance levels. For BTC, Regime~0 is the low-volatility regime ($\widehat{\sigma^2}=2.46$) and Regime~1 the high-volatility regime ($\widehat{\sigma^2}=28.61$). For ETH, Regime~0 is high volatility ($\widehat{\sigma^2}=51.83$) and Regime~1 low volatility ($\widehat{\sigma^2}=6.06$).
The autoregressive terms for both BTC and ETH show several significant coefficients, with ETH's rates of return showing more significant coefficients for high volatility regime, indicating stronger autoregressive behavior. With respect to persistency, for BTC the probability of remaining in the low-volatility regime is $\widehat{p}(0\!\to\!0)=0.771$, while for ETH the probability of remaining in the high-volatility state is $\widehat{p}(0\!\to\!0)=0.750$. The corresponding switch probabilities are moderate, for BTC, $\widehat{p}(1\!\to\!0)=0.366$ (and $\widehat{p}(0\!\to\!1)=0.229$) and, for ETH, $\widehat{p}(1\!\to\!0)=0.113$ (and $\widehat{p}(0\!\to\!1)=0.250$). These magnitudes are consistent with prior evidence of regime persistence in BTC volatility \citep{Ardia2019}. 
For ETH, several autoregressive coefficients are significant in the high-volatility regime, and the model implies a high day-to-day probability of remaining in that regime, indicating strong persistence. The estimated probability of leaving the high-volatility regime is low for ETH, suggesting high-volatility episodes are more persistent for ETH than for BTC. Overall, once a regime is entered, returns tend to remain in that regime over short horizons.

\begin{table}[htbp]
  \centering
  \caption{Markov switching model results}
  \label{TAB:markov}
  \begin{tabular}{lcccc}
    \toprule
    & \multicolumn{2}{c}{BTC} & \multicolumn{2}{c}{ETH} \\
    \cmidrule(lr){2-3} \cmidrule(lr){4-5}
    AIC & \multicolumn{2}{c}{10199.1912} & \multicolumn{2}{c}{11206.6259} \\
    BIC & \multicolumn{2}{c}{10322.3224} & \multicolumn{2}{c}{11352.1183} \\
    \cmidrule(lr){2-3} \cmidrule(lr){4-5}
          & Regime 0 & Regime 1 & Regime 0 & Regime 1 \\
    \midrule
    Constant     & 0.1208         & 0.0712         & 0.5293$^{*}$     & 0.0924 \\
                 & (0.0823)       & (0.1870)       & (0.3153)         & (0.1378) \\
    $\sigma^{2}$ & 2.4550$^{***}$ & 28.6134$^{***}$& 51.8320$^{***}$  & 6.0647$^{***}$ \\
                 & (0.7101)       & (7.7156)       & (12.8606)        & (1.4789) \\
    ar.L1        & -0.1311$^{***}$& -0.0562        & -0.1359$^{**}$   & -0.1114$^{***}$ \\
                 & (0.0374)       & (0.0742)       & (0.0543)         & (0.0397) \\
    ar.L2        & 0.0421         & 0.0072         & 0.1297$^{**}$    & -0.0398 \\
                 & (0.0468)       & (0.0728)       & (0.0659)         & (0.0423) \\
    ar.L3        & 0.0539         & -0.0301        & 0.0821           & -0.0209 \\
                 & (0.0459)       & (0.0940)       & (0.1115)         & (0.0402) \\
    ar.L4        & -0.0794$^{**}$ & 0.1206$^{***}$ & 0.0325           & -0.0142 \\
                 & (0.0368)       & (0.0381)       & (0.2085)         & (0.0759) \\
    ar.L5        & 0.0203         & -0.0581$^{**}$ & -0.0770          & 0.0028 \\
                 & (0.0325)       & (0.0294)       & (0.0959)         & (0.0566) \\
    ar.L6        & -0.0044        & -0.0515        & 0.0807           & -0.0381 \\
                 & (0.0389)       & (0.0826)       & (0.0717)         & (0.0363) \\
    ar.L7        & -0.0509        & 0.0873         & 0.0213           & -0.0071 \\
                 & (0.0433)       & (0.1027)       & (0.1005)         & (0.0442) \\
    ar.L8        & -0.0065        & -0.0169        & -0.1571$^{*}$    & 0.0199 \\
                 & (0.0318)       & (0.0278)       & (0.0917)         & (0.0304) \\
    ar.L9        &                &                & -0.0145          & -0.0203 \\
                 &                &                & (0.0385)         & (0.0344) \\
    ar.L10       &                &                & 0.1617$^{*}$     & -0.0151 \\
                 &                &                & (0.0855)         & (0.0601) \\
    \midrule
    \multicolumn{5}{c}{Regime transition parameters} \\
    \midrule
    $p(0 \to 0)$ & \multicolumn{2}{c}{0.7710$^{***}$} & \multicolumn{2}{c}{0.7502$^{***}$} \\
                 & \multicolumn{2}{c}{(0.0556)}      & \multicolumn{2}{c}{(0.0960)} \\
    $p(1 \to 0)$ & \multicolumn{2}{c}{0.3659$^{***}$} & \multicolumn{2}{c}{0.1127$^{*}$} \\
                 & \multicolumn{2}{c}{(0.1052)}      & \multicolumn{2}{c}{(0.0595)} \\
    \bottomrule
  \end{tabular}
\end{table}

\begin{figure}[htbp]
	\centering
	\caption[Volatility regimes - Bitcoin]{Volatility regimes - BTC}
			\includegraphics[width=\textwidth]{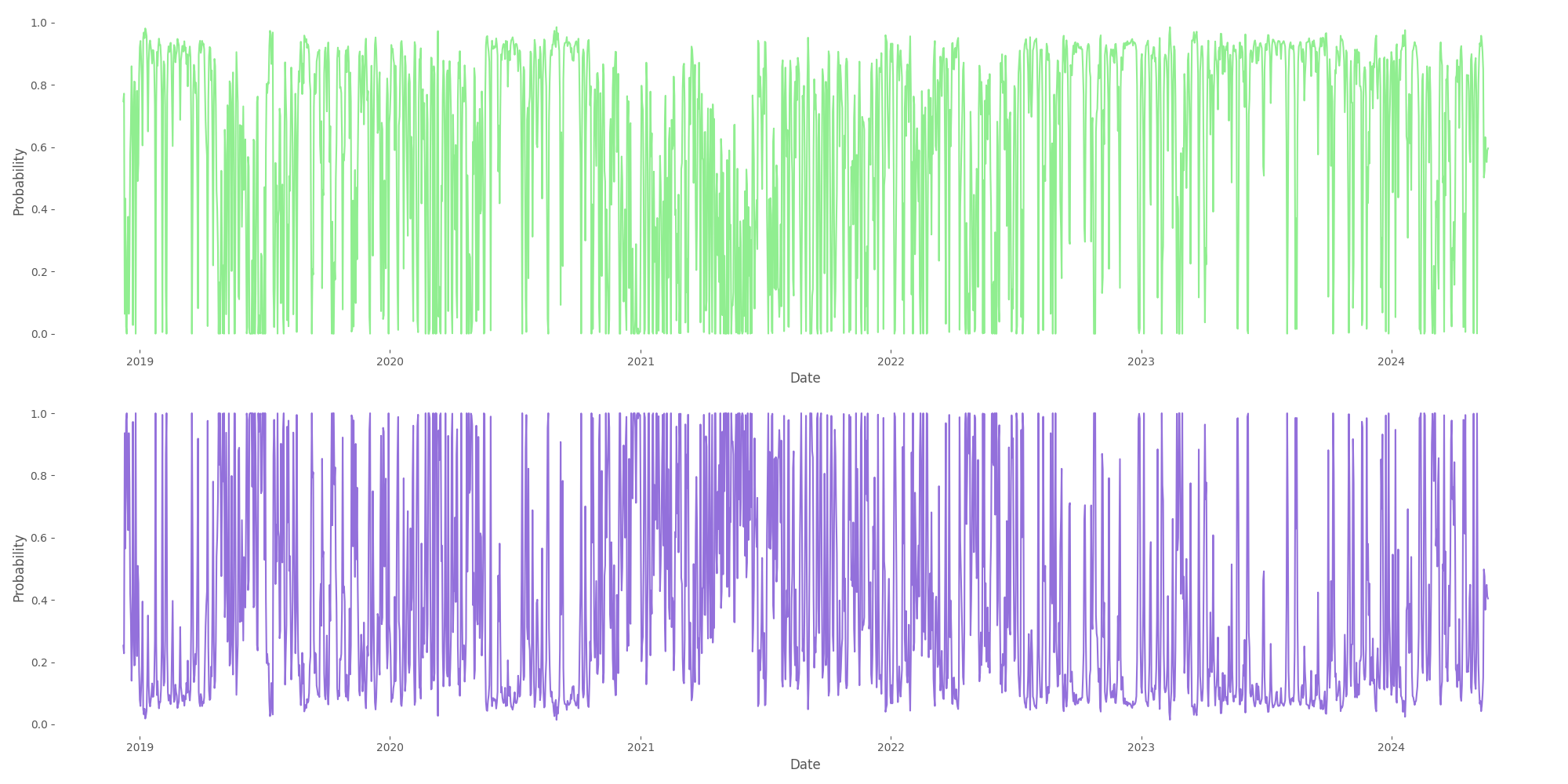}
	\label{FIG:regimes_btc}
\end{figure}

\begin{figure}[htbp]
	\centering
	\caption[Volatility regimes]{Volatility regimes - ETH}
			\includegraphics[width=\textwidth]{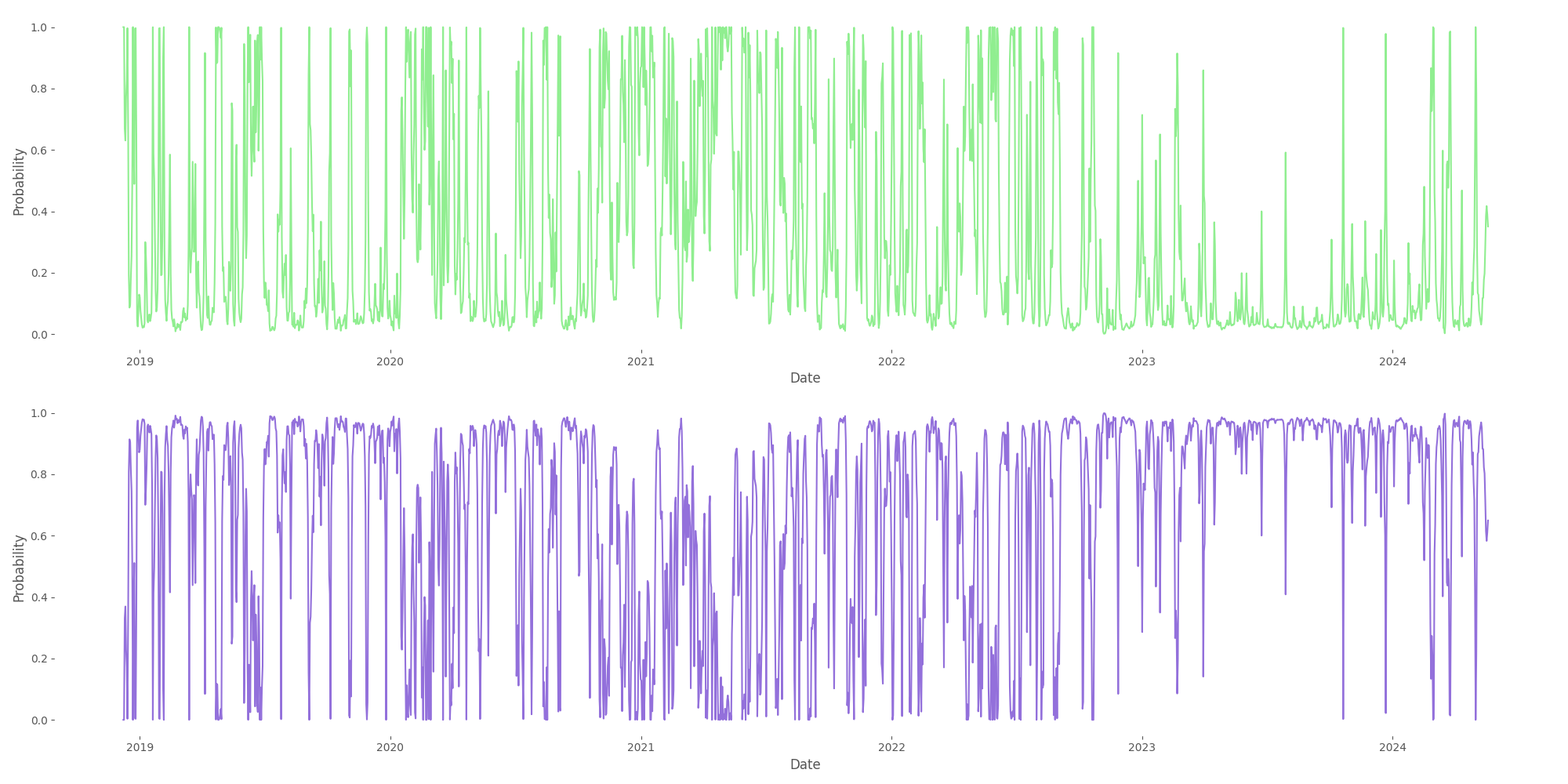}
	\label{FIG:regimes_eth}
\end{figure}

Figures \ref{FIG:regimes_btc} and \ref{FIG:regimes_eth} depict smoothed probabilities of regime 0 (low volatility for BTC and high volatility for ETH) in green and regime 1 (high volatility for BTC and low volatility for ETH) in purple from the MS-AR model. For BTC, the low volatility regime shows more frequent transitions between low and high volatility states. The interruption in persistence staying in a low volatility regime is observable around early 2020, i.e., the bear market, and late  2020 to early 2021, i.e., the bull market. The high volatility regime shows significant periods around these dates as well, highlighting market reactions. For ETH, both regimes show less frequent transitions and show more pronounced peaks, indicating increased market activity. On the contrary, the low volatility regime shows persistence after late 2021, which mirrors the growing adoption and the establishment of the ETH chain in the decentralized finance ecosystem. 
Both BTC and ETH show significant reactions to global market events, pinned down by frequent regime changes and definite peaks, correspondingly.

Table~\ref{TAB:garch} reports the regime–specific volatility models selected after the MS–AR classification, where the model choice is based on the BIC. 
Standard errors (in parentheses) of estimated coefficients for the mean (top) and variance (middle) equations and the error distribution (bottom) are robust to heteroskedasticity and autocorrelation. The significance is based on $t$-statistics: $^{***}$, $^{**}$, $^{*}$ denote the 1\%, 5\%, and 10\% levels. The GJR term $\gamma_1$ appears only in the GJR--GARCH specification. Goodness-of-fit is assessed with Ljung--Box and Engle's ARCH diagnostics. Only for BTC’s low–volatility regime the optimal process is a GJR–GARCH(1,1,1). The estimated leverage parameter is negative and highly significant, indicating an inverse leverage effect (positive shocks raise volatility more than negative shocks). This pattern, unusual for equities but repeatedly documented for BTC, aligns with findings in \citet{Bouri2017}, \citet{Katsiampa2017}, \citet{Stavroyiannis2018}, and \citet{Ardia2019}. On the contrary, the remaining regimes (BTC high–volatility, ETH high– and low–volatility) are best described by standard GARCH(1,1), which is the GJR model with a zero leverage term \( \gamma \). Across regimes, the \( \beta \) coefficients are large and significant, confirming strong volatility persistence. They are somewhat lower in high–volatility states, implying faster mean reversion when markets are stressed. For ETH, persistence in the low–volatility state is very high (large \( \beta \)), while the high–volatility state shows persistence that is lower than BTC’s, suggesting milder clustering in ETH during turbulent periods. Diagnostic checks support the adequacy of the fitted models: across all regimes, the Ljung–Box test on standardized residuals yields p-values above 0.10 (0.1608, 0.9965, 0.9999, 0.7628; Table~\ref{TAB:garch}), so we fail to reject the null of no residual autocorrelation and Engle’s ARCH tests likewise have p-values above 0.10, indicating no remaining ARCH effects. These results are consistent with a broader literature showing that cryptocurrency volatility is regime dependent, heavy–tailed, and sensitive to breaks and jumps found in \citet{charles2019}, \citet{bouri2019}, \citet{wu2021}, \citet{venter2020}, \citet{liang2022}. Taken together, the evidence justifies our regime–specific, skewed–GARCH specification and supports using these regime–conditioned volatilities as inputs to the benchmark pricing step.

\begin{table}[htbp]
  \centering
  \caption{GARCH results}
  \label{TAB:garch}
  \begin{tabular}{lcccc}
    \toprule
    & \multicolumn{2}{c}{BTC} & \multicolumn{2}{c}{ETH} \\
    \cmidrule(lr){2-3}\cmidrule(lr){4-5}
    Regime & Regime 0 & Regime 1 & Regime 0 & Regime 1 \\
    Interpretation & Low volatility & High volatility & High volatility & Low volatility \\
    Specification       & GJR--GARCH(1,1,1) & GARCH(1,1) & GARCH(1,1) & GARCH(1,1) \\
    AIC    & 6785.3842 & 3294.1601 & 3025.3312 & 8079.4733 \\
    BIC    & 6822.0511 & 3320.5530 & 3050.7008 & 8111.2821 \\
    \midrule
    \multicolumn{5}{c}{Mean model}\\
    \midrule
    $\mu$         & 0.2107$^{**}$ & 0.1197 & 0.3629 & 0.1315 \\
                  & (0.0993)      & (0.1652) & (0.2269) & (0.0945) \\
    \midrule
    \multicolumn{5}{c}{Volatility model}\\
    \midrule
    $\omega$      & 0.1620 & 3.5636$^{*}$ & 8.1344$^{*}$ & 0.1500 \\
                  & (0.5642) & (2.0924) & (4.2226) & (0.2621) \\
    $\alpha_1$    & 0.1141 & 0.1336$^{*}$ & 0.1325$^{*}$ & 0.0674$^{*}$ \\
                  & (0.0714) & (0.0591) & (0.0736) & (0.0397) \\
    $\gamma_1$    & -0.0895$^{***}$ &  &  &  \\
                  & (0.0255)        &  &  &  \\
    $\beta_1$     & 0.9306$^{***}$ & 0.7304$^{***}$ & 0.6330$^{***}$ & 0.9325$^{***}$ \\
                  & (0.1161) & (0.1029) & (0.1429) & (0.0485) \\
    \midrule
    \multicolumn{5}{c}{Distribution}\\
    \midrule
    $\eta$        & 2.9555$^{***}$ & 2.9625$^{***}$ & 3.2018$^{***}$ & 3.5463$^{***}$ \\
                  & (0.4766) & (0.3934) & (0.5555) & (0.3412) \\
    $\lambda$     & 0.0602 & 0.0054 & 0.0044 & 0.0277 \\
                  & (0.0467) & (0.0496) & (0.0539) & (0.0333) \\
    \midrule
    \multicolumn{5}{c}{Diagnostics}\\
    \midrule
    Ljung--Box  & 14.2746 & 1.9784 & 0.4306 & 6.5972 \\
    $p$-value             & 0.1608  & 0.9965 & 0.9999 & 0.7628 \\
    Engle's ARCH & 0.8115 & 0.0317 & 0.0287 & 1.3423 \\
    $p$-value             & 0.9999  & 0.9999 & 0.9999 & 0.9993 \\
    \bottomrule
  \end{tabular}
\end{table}

\begin{sidewaysfigure}[htbp]
	\centering
	\caption[Volatility dynamics]{Volatility dynamics - BTC}
			\includegraphics[width=1.10\textwidth]{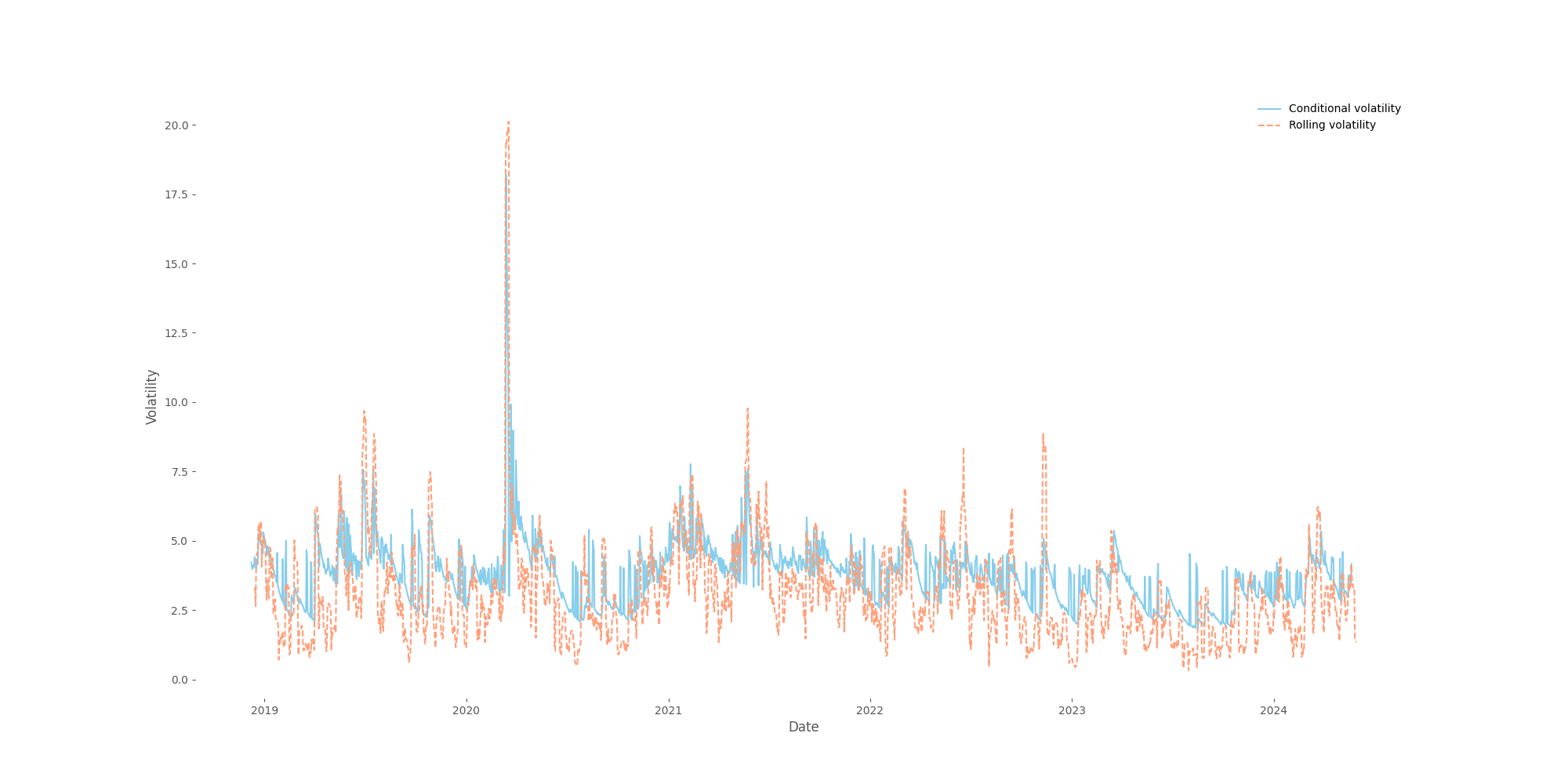}
	\label{FIG:vols_BTC}
\end{sidewaysfigure}

\begin{sidewaysfigure}[htbp]
	\centering
	\caption[Volatility dynamics]{Volatility dynamics - ETH}
			\includegraphics[width=1.10\textwidth]{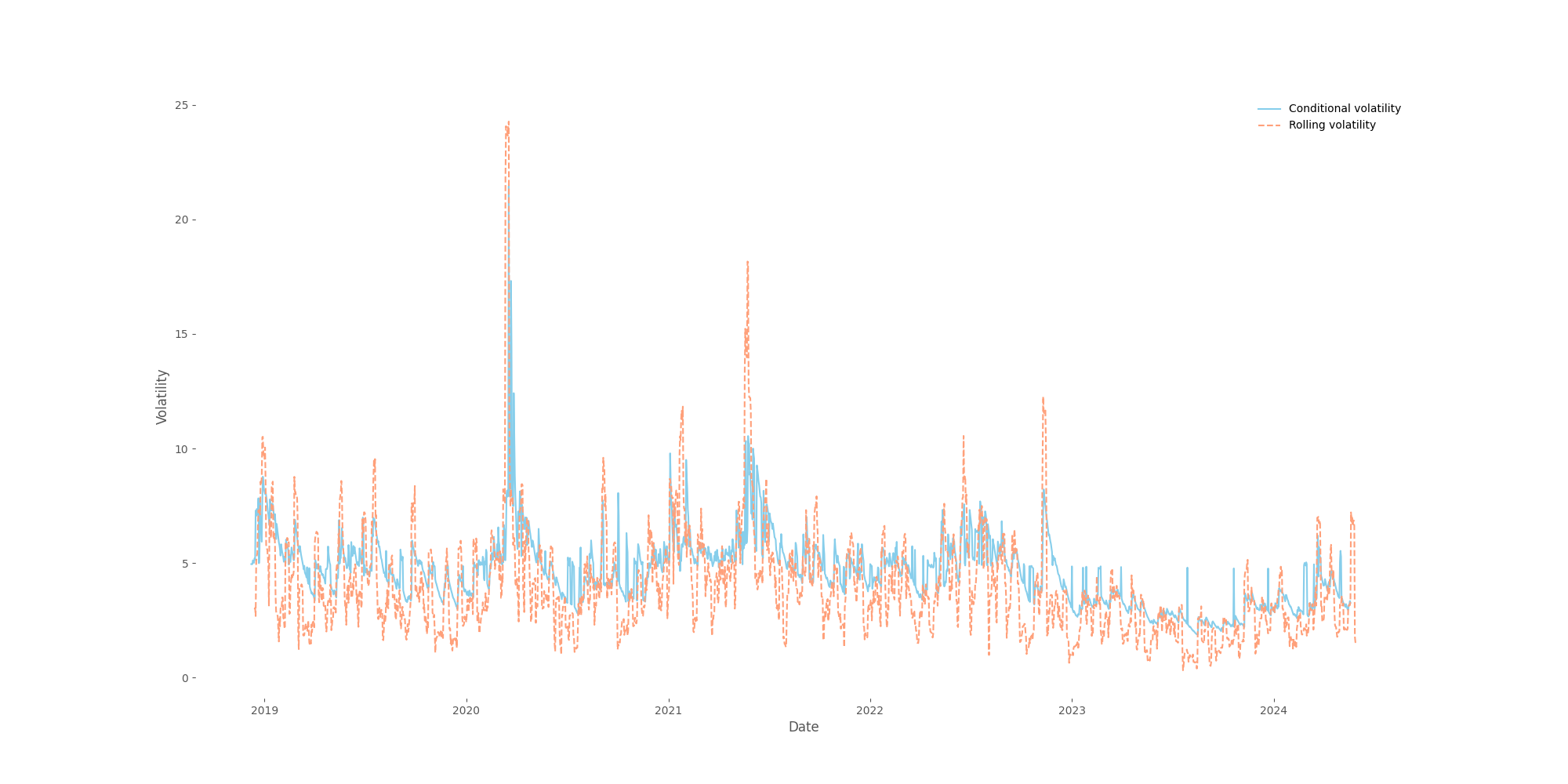}
	\label{FIG:vols_ETH}
\end{sidewaysfigure}

Figures \ref{FIG:vols_BTC} and \ref{FIG:vols_ETH} compare two volatility measures for BTC and ETH: the model–implied conditional volatility from the MS–AR–(GJR)–GARCH specification (blue) and a 7-day rolling realized volatility (orange). The rolling series is intentionally very reactive and, therefore, noisy from day to day. On the contrary, the regime–sensitive GARCH path is smoother in normal periods but shows sharper spikes around major market moves. This pattern is expected when volatility is clustered, exhibits regime shifts, and contains jump components. Short windows pick up the transient noise, while regime–dependent models filter that noise yet still respond strongly when the regime changes \citep{Thies2018,Ardia2019,charles2019,bouri2019,wu2021}. 
In our sample, the ETH path shows taller peaks than BTC, indicating stronger short–run reactions during turbulent intervals. While our setting differs, \citet{AndolfattoNaikSchoenleber2024} likewise document rapid movements in on–chain IV and link them to order–flow pressures on ETH. Overall, the two volatility measures co–move closely and agree on the timing of volatility bursts, but the regime–switching model reduces high–frequency noise and highlights the major episodes, providing a cleaner volatility estimate for benchmark pricing.

Table~\ref{TAB:statsmispricing} reports summary statistics for mispricing \(\Delta\text{price}_{j}\), defined in equation \ref{eq:misprice}. Because Hegic allows fractional purchases, premiums are first expressed on a per-contract basis so that the ratio is scale-free. Mean mispricing is positive in all samples except for ETH put options, implying that, on average, the BS reference exceeds the Hegic quote (it is lower for ETH put options). Mean levels are smaller for ETH than for wBTC, indicating closer alignment of ETH quotes with the benchmark price. Dispersion is higher for wBTC, especially for call options, as shown by the larger standard deviations. The minima and maxima document occasional large deviations in both directions for both underlyings. Quartiles suggest slightly wider spread for wBTC than for ETH. Distributions are right-skewed, consistent with more observations where the benchmark price is above the Hegic quote. This asymmetry is more pronounced for call options. High kurtosis further indicates fat tails, i.e., infrequent but sizable dislocations, again most evident for call options.

\begin{sidewaysfigure}[htbp]
	\centering
	\caption{Mispricing - wBTC call options}
	\label{FIG:mispricing_call_BTC}
	\includegraphics[width=\textwidth]{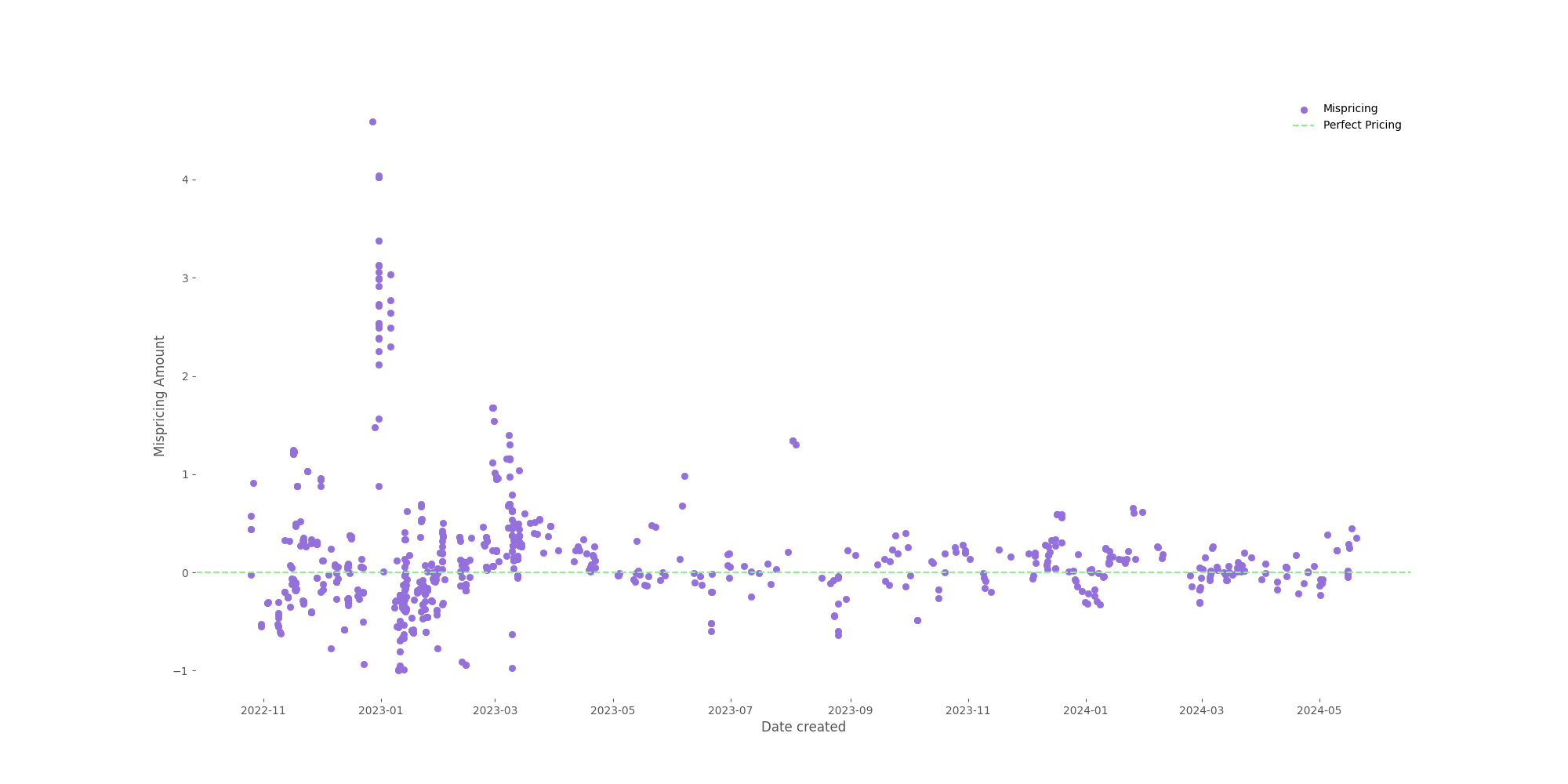}
\end{sidewaysfigure}

\begin{sidewaysfigure}[htbp]
	\centering
	\caption{Mispricing - wBTC put options}
	\label{FIG:mispricing_put_BTC}
	\includegraphics[width=\textwidth]{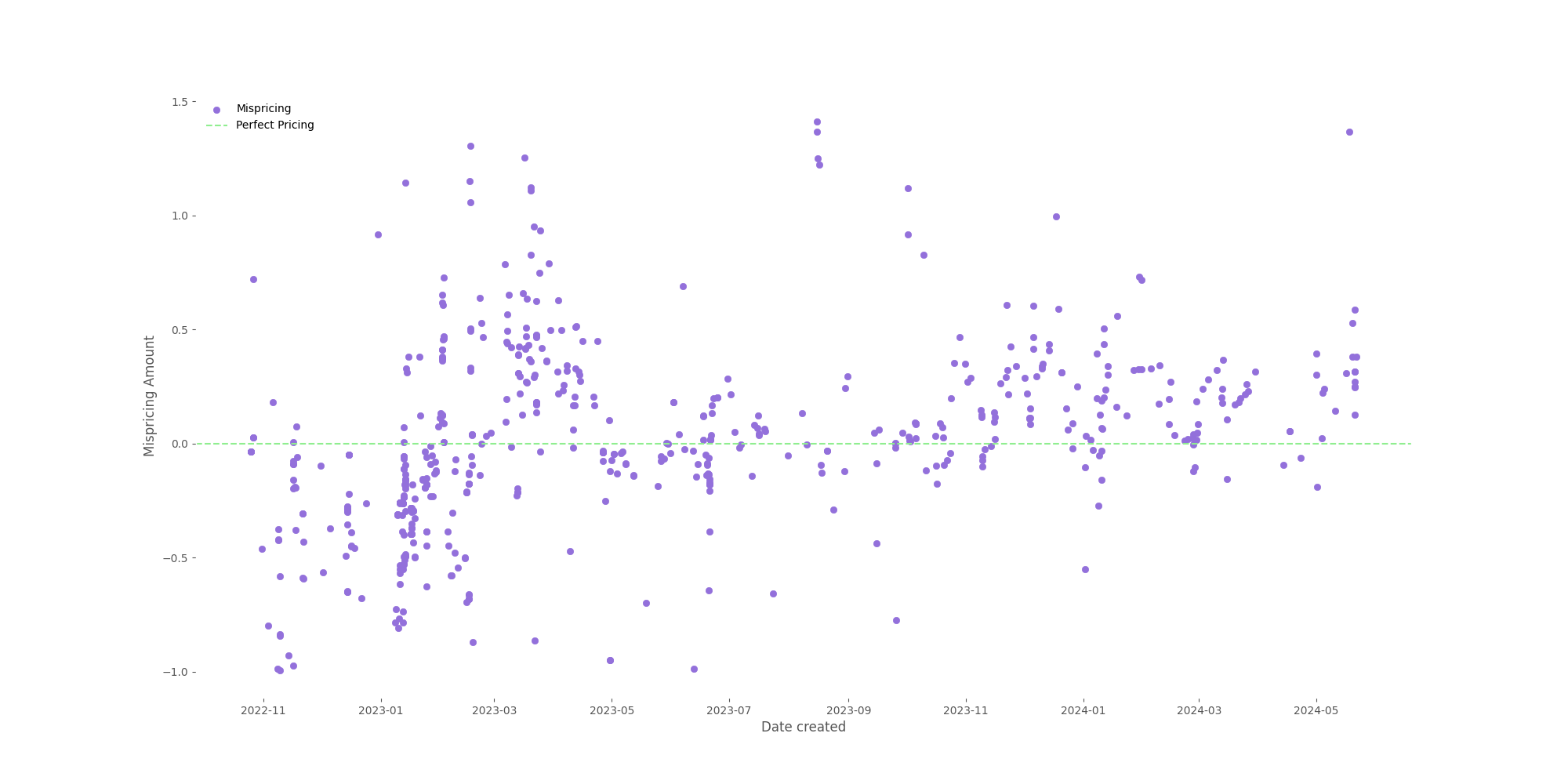}
\end{sidewaysfigure}

\begin{sidewaysfigure}[htbp]
	\centering
	\caption{Mispricing - ETH call options}
	\label{FIG:mispricing_call_ETH}
	\includegraphics[width=\textwidth]{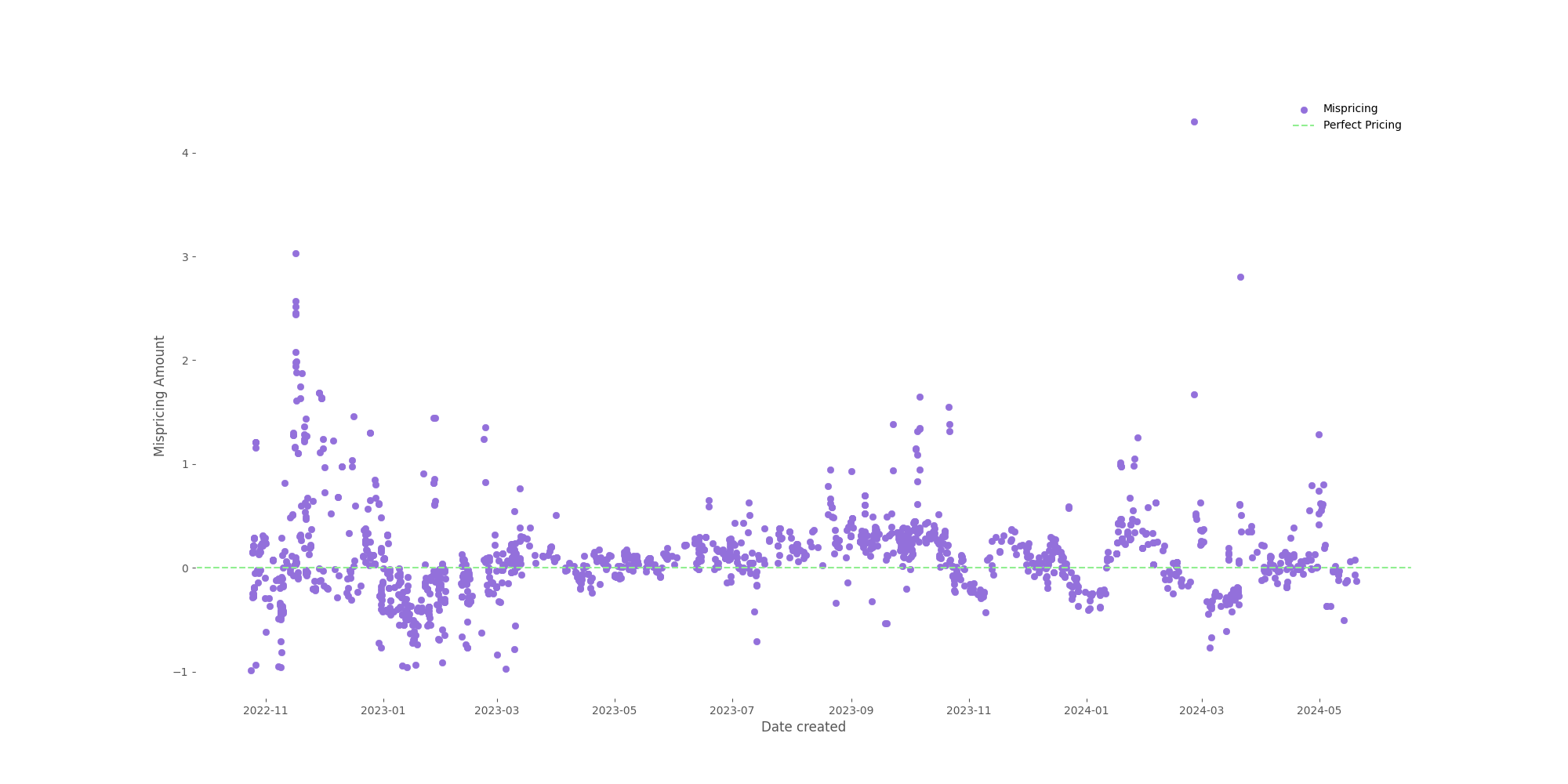}
\end{sidewaysfigure}

\begin{sidewaysfigure}[htbp]
	\centering
	\caption{Mispricing - ETH put options}
	\label{FIG:mispricing_put_ETH}
	\includegraphics[width=\textwidth]{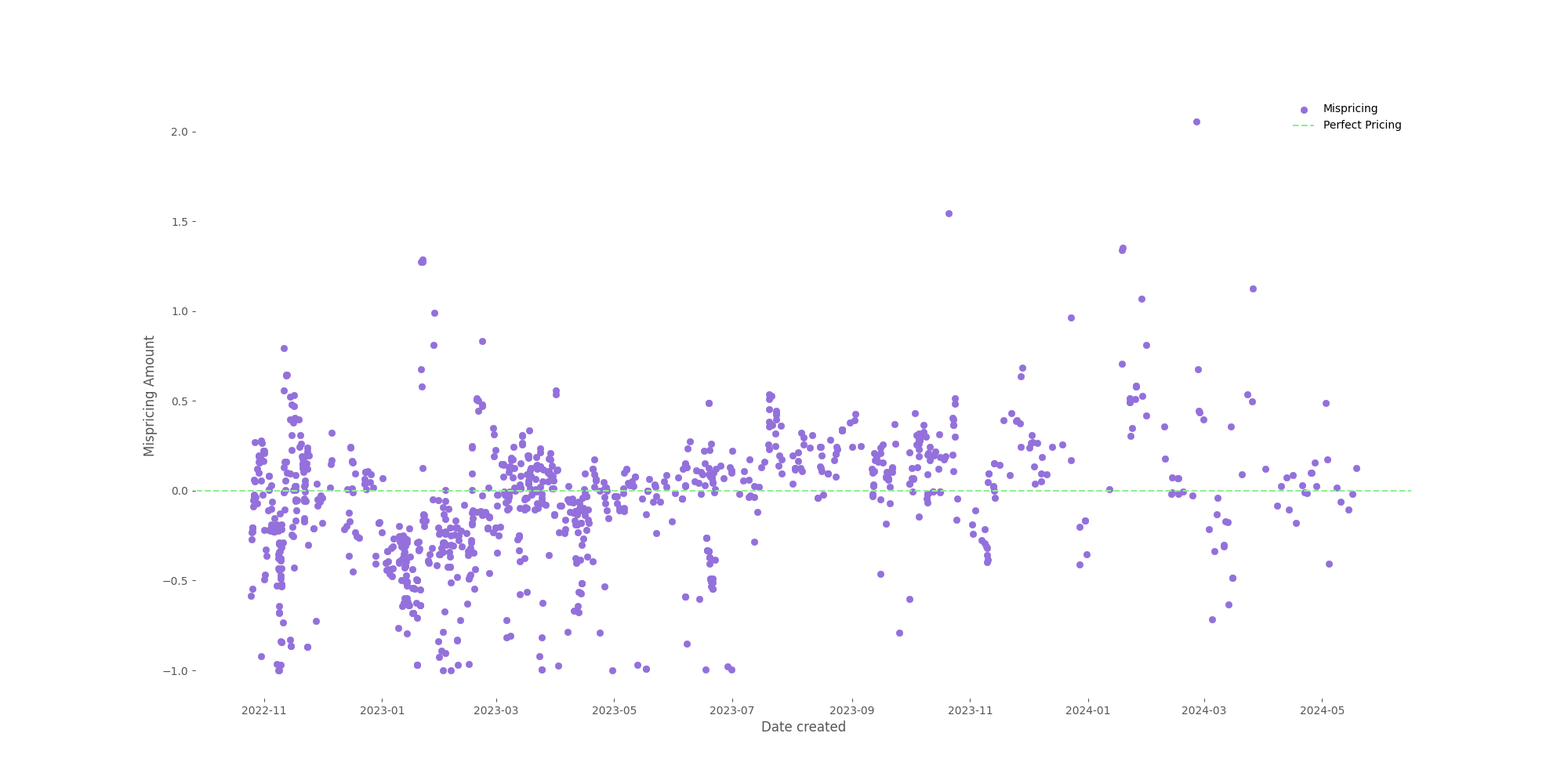}
\end{sidewaysfigure}

Before estimating the FGLS model, we diagnose and preprocess the regressors to avoid spurious conclusion and to make effect sizes comparable across variables. First, we compute a pairwise correlation matrix (figure \ref{FIG:correl}) and find no severe correlations. For wBTC, most pairs are low with a few moderate entries, e.g., Strike with Type) and, for ETH, moderate correlations appear for Volume with Volatility, which is consistent with higher activity during volatile periods. Second, we assess multicollinearity using variance inflation factors (table \ref{TAB:vif}), where all VIFs lie below 5, indicating that collinearity is not a concern. Third, to reduce right-skewness and stabilize scale, we take natural logs of strictly non-negative variables, i.e., Amount, Strike, Maturity, and Volume. Finally, to place coefficients on a comparable footing, we standardize all continuous regressors (except dummy variables) to zero mean and unit variance. These transformations are standard in empirical asset-pricing and market-microstructure applications, improving efficiency and interpretability in heteroskedastic cross-sections.\footnote{See \citet[pp.~204–210]{CampbellLoMacKinlay1998}, \citet[Ch.~6]{Wooldridge2010}.}

Table \ref{TAB:FGLSresults} reports the FGLS estimates for wBTC and ETH options with Newey–West (HAC) standard errors in parentheses and statistical significance with $^{***}$, $^{**}$, $^{*}$, denoting the \SI{1}{\%}, \SI{5}{\%}, and \SI{10}{\%} levels, respectively. Adjusted $R^2$, the model $F$-statistic, and the Wald statistic are reported as goodness-of-fit measures.
Recall that continuous regressors (Amount, Strike, Maturity, Return, Volume, Volatility) are logs and standardized to zero mean and unit variance, so each slope can be read as the change in mispricing (in \%age points) for a one–standard–deviation move in the regressor. Model fit is at the comparable level across underlyings with Adj.~$R^2{=}0.56$ for wBTC and $0.50$ for ETH. Intercepts are negative but statistically insignificant in both markets, indicating no detectable average bias.

For wBTC, Amount, Strike, Maturity, Volume, Volatility, and the Kind dummy are statistically significant. Translating coefficients into US dollars using the mean wBTC call premium ($\approx\$700$ according to table \ref{tab:DescriptiveStats}) yields:  
(i) Order size coefficient of \(0.0228^{***}\) implies that a $1$ s.d. increase in Amount raises the mispricing by about \(2.28\%\), i.e., roughly \(\$16\) on a \$700 option.  
(ii) Strike coefficient of \(0.0559^{***}\) adds \(5.59\%\), or about \(\$39\), consistent with wider prices deviation for further OTM options.  
(iii) Maturity coefficient of  \(0.0166^{*}\) adds \(1.66\%\), or about \(\$12\), indicating slightly larger price spreads for longer maturities.  
(iv) Underlying depth (Volume) coefficient of \(-0.0743^{***}\) reduces the price difference by \(7.43\%\), or about \(-\$52\), consistent with tighter alignment when market activity is higher.  
(v) Volatility coefficient of \(0.0789^{***}\) increases the price deviation by \(7.89\%\), or about \(\$55\), in line with greater pressure on inventory/oracle channels in turbulent periods.  
(vi) Option side (kind) coefficient of \(-0.0896^{***}\) indicates call options are, on average, \(8.96\%\) closer to the benchmark price than put options (about \(-\$63\) on a \$700 premium).

For ETH, only Amount, Volume, and Kind are significant, with signs matching wBTC results where comparable. Using the mean ETH call premium ($\approx\$673$) yields:  
(i) Order size coefficient of \(0.0334^{***}\) implies an increase in mispricing by \(+3.34\%\), or about \(\$22\).  
(ii) Underlying depth (Volume) coefficient of \(-0.1353^{***}\) implies decrease of\(-13.53\%\) in the price deviation, or about \(-\$92\), underscoring the role of liquidity.  
(iii) Option side (Kind) coefficient of \(0.0869^{**}\) indicates call options show about \(+8.69\%\) larger price difference than put options, i.e., roughly \(+\$59\) on a \$673 premium. Coefficients on strike, maturity, return, and volatility are not statistically different from zero in the ETH sample, therefore, and we do not ascribe economic content to them.

The pattern of larger price deviations for bigger orders, further OTM strikes, longer maturities (wBTC), and higher volatility, and smaller deviations when underlying volume is high, is consistent with pooled–liquidity AMM mechanics and oracle frictions documented in prior work \citep{alexander2023,eskandari2017,priem2022,volkovich2023}.
In practical terms, the estimated effects suggest following calibration opportunities for the protocol. First, introduce an amount slope in the rate schedule so that larger orders pay more, compensating the pool for convexity and inventory pressure. Second, raise OTM rates relative to ATM to reflect greater replication difficulty \citep{madan2019,olivares2020}. Third, allow systemic adjustments to the base rate across maturities, if longer maturities increase collateral and, thus, increase funding needs under full collateralization \citep{priem2022}. 
Fourth, link fees (or small rate adjustments) to observed market depth. When liquidity depth is high, reduce fees to bring quotes closer to the benchmark and, when liquidity depth is low, keep fees higher to protect the pool from inventory risk. This aligns with evidence that liquidity conditions shape cross‐exchange price spreads \citep{kristoufek2023,volkovich2023}.

\begin{table}[htbp]
  \centering
  \caption{FGLS results}
  \label{TAB:FGLSresults}
  \begin{tabular}{lcc}
    \toprule
    & \multicolumn{1}{c}{wBTC options} & \multicolumn{1}{c}{ETH options} \\
    \midrule
    Intercept             & -0.0213        & -0.2700        \\
                          & (0.0512)       & (0.1775)       \\
    Amount                & 0.0228$^{***}$ & 0.0334$^{***}$ \\
                          & (0.0086)       & (0.0107)       \\
    Strike                & 0.0559$^{***}$ & -0.0160        \\
                          & (0.0204)       & (0.0267)       \\
    Maturity              & 0.0166$^{*}$   & -0.0087        \\
                          & (0.0094)       & (0.0118)       \\
    Return (underlying)   & 0.0005         & -0.0167        \\
                          & (0.0082)       & (0.0145)       \\
    Volume (underlying)   & -0.0743$^{***}$& -0.1353$^{***}$\\
                          & (0.0140)       & (0.0207)       \\
    Volatility (underlying)& 0.0789$^{***}$& -0.0662        \\
                          & (0.0230)       & (0.0554)       \\
    Kind                  & -0.0896$^{***}$& 0.0869$^{**}$  \\
                          & (0.0337)       & (0.0362)       \\
    Type                  & 0.0280         & 0.1827         \\
                          & (0.0424)       & (0.1681)       \\
    \midrule
    \multicolumn{3}{c}{Diagnostics}\\
    \midrule
    Adj.\ $R^{2}$         & 0.5617         & 0.4961         \\
    Cond.\ number         & 46.5044        & 38.8960        \\
    F-statistic           & 39.1036        & 28.5514        \\
    $p$-value             & 0.0000         & 0.0000         \\
    Wald-statistic        & 312.8189       & 228.3650       \\
    $p$-value             & 0.0000         & 0.0000         \\
    \bottomrule
  \end{tabular}

\end{table}

A key driver of differences between our benchmark price and Hegic quotes is the volatility input used in the pricing. In practice, IV is the standard diagnostic in risk management, forecasting, and option valuation. Accordingly, for each Hegic trade we invert the \citet{BlackScholes1973} formula with a Brent root--finding to estimate the option’s IV and compare it to the volatility $\sigma_t$ from our MS--AR--(GJR)--GARCH model. If $\text{IV}>\sigma_t$, the Hegic quote embeds a volatility premium and, if $\text{IV}<\sigma_t$, it embeds a discount. However, this comparison is not proof of arbitrage, because execution costs, limited depth, and oracle timing can block riskless trades. The possible trading strategies could follow as: when $\text{IV}<\sigma_t$ (the option is underpriced relative to our reference), a trader can buy the option on Hegic and form a delta--neutral hedge by shorting $\Delta$ units of the underlying (at spot or futures markets). This isolates the volatility exposure. When $\text{IV}>\sigma_t$ (the option is overpriced), the mirror trade is to sell the option and hedge delta. Because Hegic does not allow users to write options, that short leg must be executed off-protocol, or indirectly via liquidity provision into the pool. Related work demonstrates the feasibility of volatility-premium strategies on protocols with two-sided markets \citep{AndolfattoNaikSchoenleber2024}. Frictions in DeFi, e.g., fees, slippage, depth, and oracle update cadence, constrain realized profits in both directions. The IV--$\sigma_t$ signal also has clear implications for liquidity providers. If quotes carry a sustained premium ($\text{IV}>\sigma_t$) and realized volatility remains moderate, filled trades tend to deliver higher premia with fewer exercises, benefiting the pool. If quotes carry a discount ($\text{IV}<\sigma_t$), the pool collects too little premium for the convexity it sells and is more exposed when volatility rises. 

\begin{sidewaysfigure}[htbp]
	\centering
	\caption{Volatility difference - wBTC call options}
	\label{FIG:voldiff_call_BTC}
	\includegraphics[width=\textwidth]{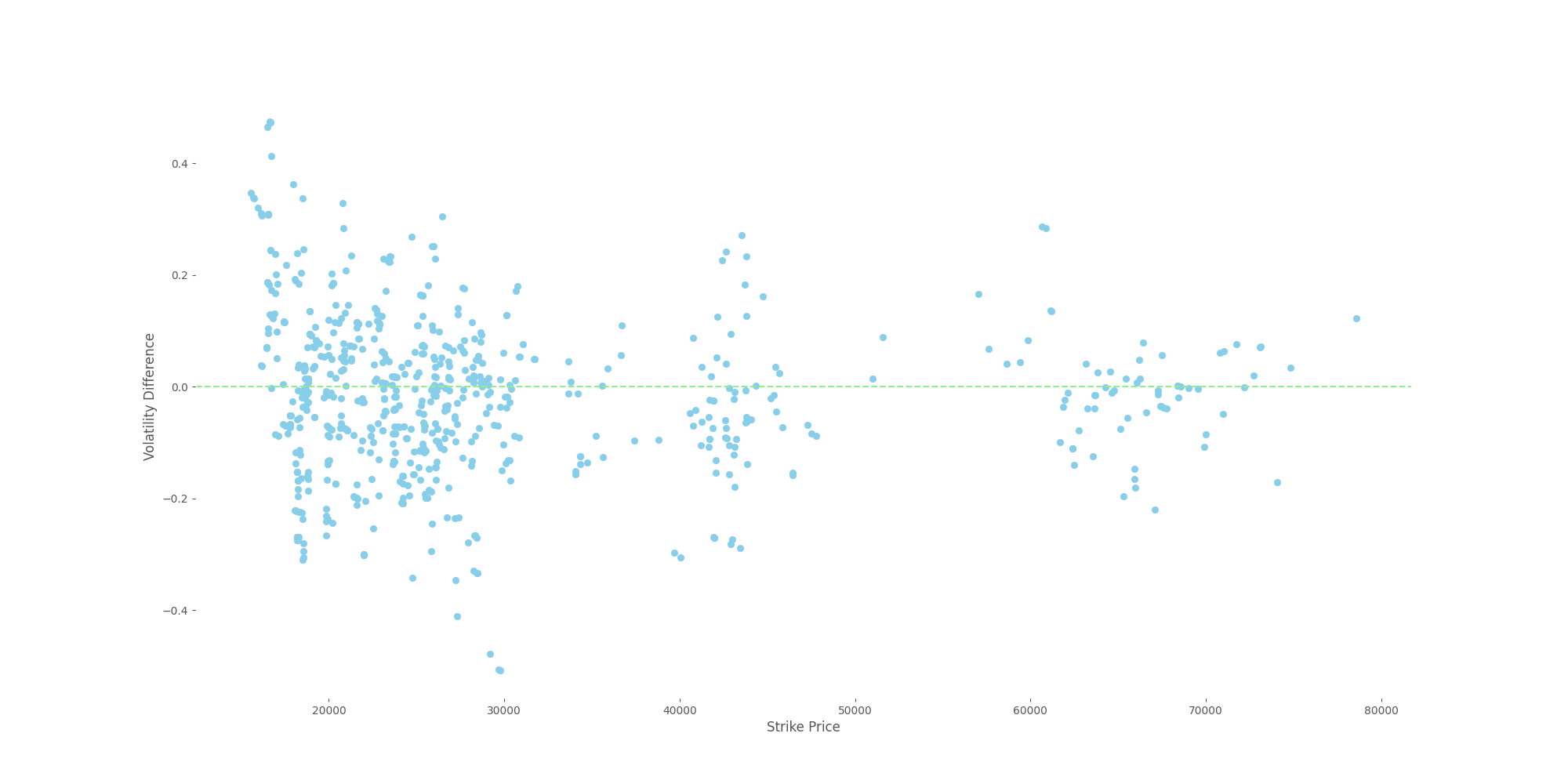}
\end{sidewaysfigure}

\begin{sidewaysfigure}[htbp]
	\centering
	\caption{Volatility difference - wBTC put options}
	\label{FIG:voldiff_put_BTC}
	\includegraphics[width=\textwidth]{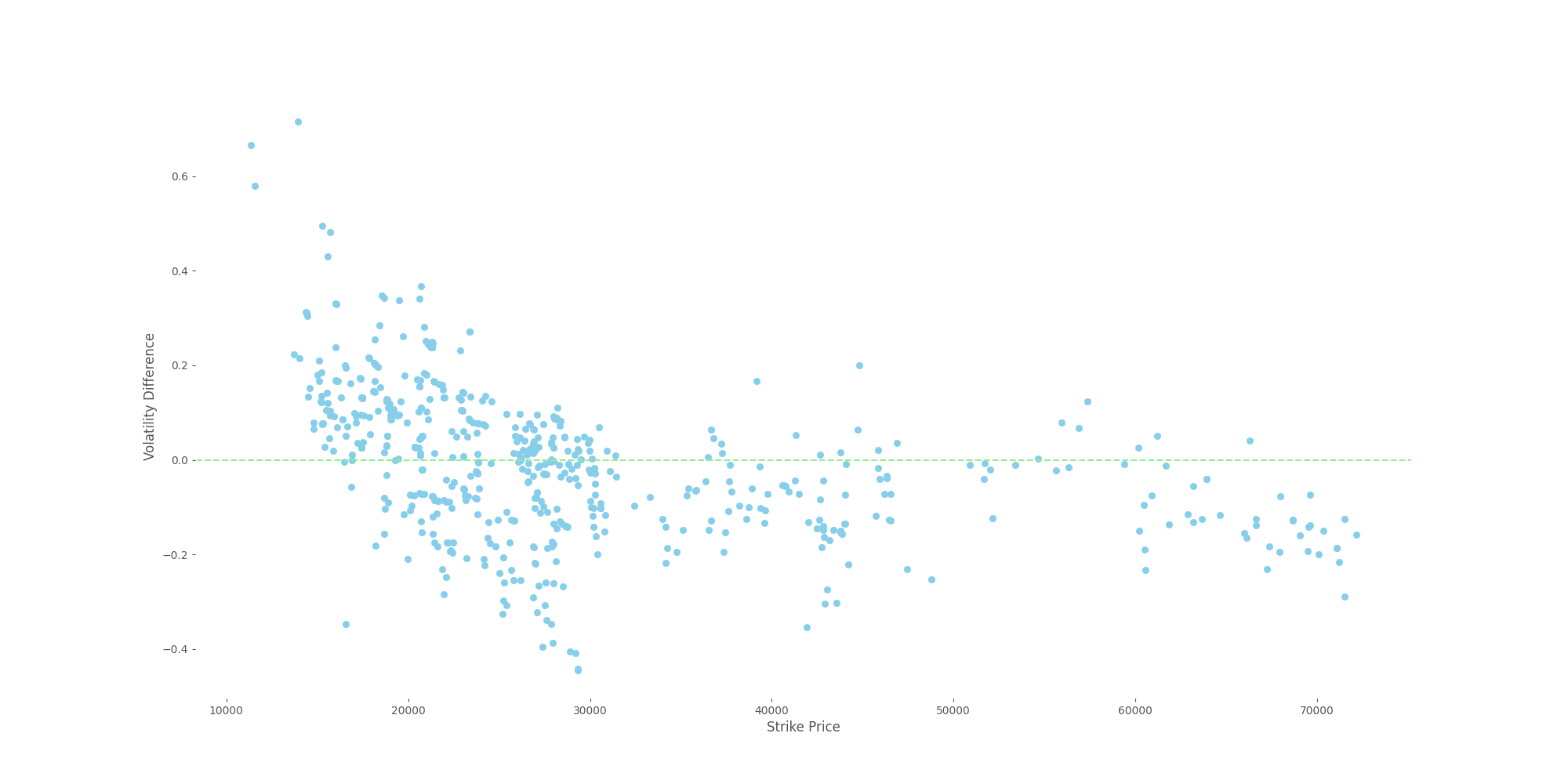}
\end{sidewaysfigure}

\begin{sidewaysfigure}[htbp]
	\centering
	\caption{Volatility difference - ETH call options}
	\label{FIG:voldiff_call_ETH}
	\includegraphics[width=\textwidth]{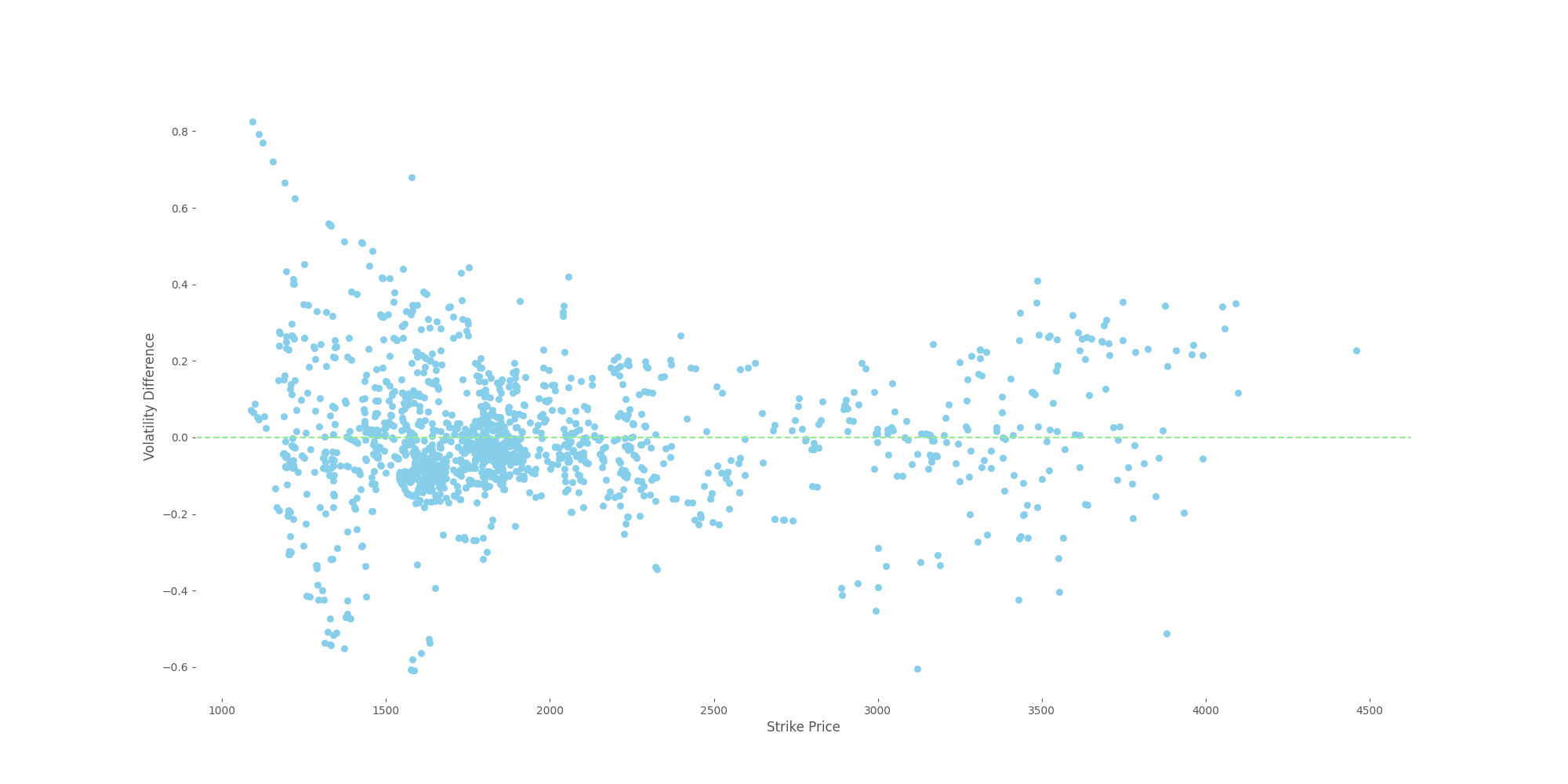}
\end{sidewaysfigure}

\begin{sidewaysfigure}[htbp]
	\centering
	\caption{Volatility difference - ETH put options}
	\label{FIG:voldiff_put_ETH}
	\includegraphics[width=\textwidth]{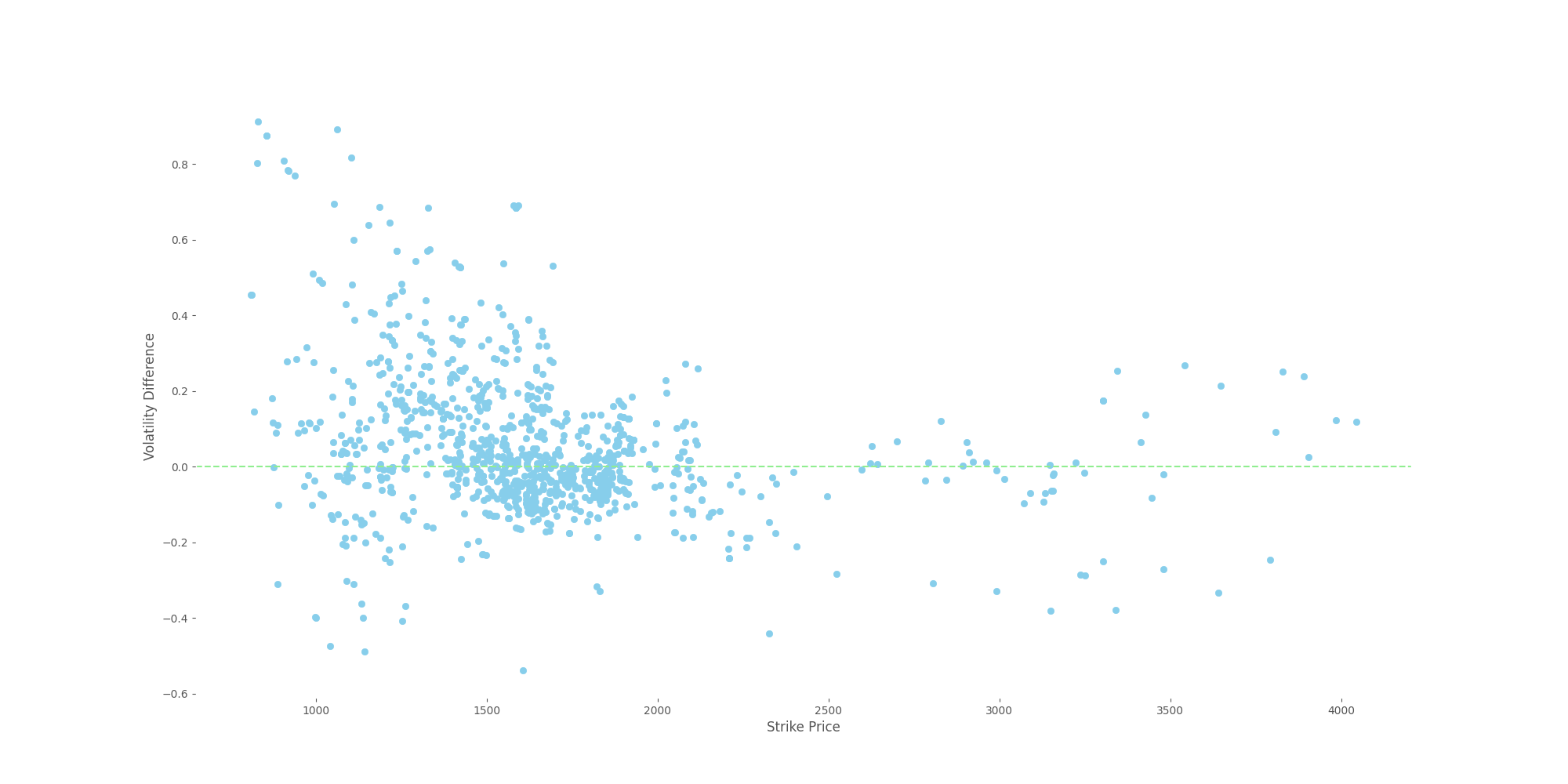}
\end{sidewaysfigure}

In the BS benchmark, we treat volatility estimate as fixed over the life of the option. For Hegic’s short maturities (7–90 days), this is a practical approximation. However, for longer maturities, the volatility estimate can drift if the market switches regimes after trade time. To keep track of this risk, the smoothed probabilities of being in (or moving into) the high–volatility regime can help to identify when a fixed–volatility assumption is more likely to understate future volatiltiy \citep{Thies2018,Ardia2019,charles2019,bouri2019,wu2021}. For each Hegic quote, we invert BS to estimate its IV. Figures \ref{FIG:voldiff_call_BTC}, \ref{FIG:mispricing_put_BTC}, \ref{FIG:mispricing_call_ETH} and \ref{FIG:mispricing_put_ETH} plot \(\text{IV}-\sigma_t\) by strike, and table \ref{TAB:statsvoldifff} reports its summary statistics. For wBTC, call option differences have a slightly negative mean (about \(-0.02\)), so IV is, on average, a bit below \(\sigma_t\); put options are also slightly negative and very close to zero (tighter alignment). For ETH, call option differences are near zero on average (slightly negative), while put option differences are positive on average (about \(+0.05\)) with occasional large positive outliers (maximum around \(0.91\)). Dispersion and tails are larger for ETH (especially on puts) consistent with stronger sensitivity to market spikes and with evidence that on–chain ETH option metrics respond more to changing conditions \citep{AndolfattoNaikSchoenleber2024}.

These \(\text{IV}-\sigma_t\) diagnostics indicate where a delta–neutral long– (if \(\text{IV}<\sigma_t\)) or short–volatility strategy (if \(\text{IV}>\sigma_t\)) would be possible, subject to the practical limits noted above. 
A delta–neutral position removes first–order exposure to the underlying and turns the trade into a bet on the spread between the volatility embedded in the option’s price (IV) and the volatility the market will actually realize over the life of the contract. In continuous time, the P\&L of a delta–hedged option decomposes into gains from gamma (which scale with realized variance) minus theta (time decay), plus trading frictions. Therefore, a long option with frequent hedging earns on average if realized volatility exceeds the level implied by IV, and loses if realized variance is lower \citep[chap.~20]{Hull2012}. In our setting, when \(\text{IV}<\sigma_t\), buying the option on Hegic and delta–hedging in spot/futures is expected to have positive gain. When \(\text{IV}>\sigma_t\), the opposite trade would have positive gain, though that short leg must be placed off–protocol or approximated via liquidity provision. The size of any premium is bounded by discrete hedging error, funding costs for the hedge, exchange fees, slippage, and oracle/update latency. These frictions rise in stressed markets and can overturn profits even when the IV–\(\sigma_t\) signal is favorable \citep{akyildirim2020,sebastiano2020,volkovich2023}.

\section{Conclusion}\label{sec:conclusion}

This paper studies on–chain option pricing on Hegic using transaction–level data for wBTC and ETH on Arbitrum chain (October 24, 2022 to May 21, 2024) and a BS reference price with a regime–sensitive volatility. Three results are robust across specifications. First, mispricing co–moves positively with order size and negatively with underlying trading volume in both markets. For wBTC it also increases with strike, maturity, and volatility, while the corresponding ETH coefficients are statistically indistinguishable from zero. Second, the kind (call vs.\ put) effect is significant with opposite signs across underlyings, indicating side–specific pressure that is protocol– and state–dependent. 
The cross–section analysis offers calibration levers that preserve Hegic’s rate–based quoting while improving price quality. (i) Introduce a size–dependent increase so that larger trades pay a higher effective rate, compensating pooled writers for inventory and convexity pressure. (ii) Raise rates for OTM strikes relative to ATM to reflect tougher replication and thinner markets\citep{madan2019,olivares2020}. (iii) Apply rate adjustment for longer maturities to acknowledge capital being locked up longer under full collateralization \citep{priem2022}. (iv) Condition fees on observed market liquidity depth so that deviations narrow when liquidity is high \citep{kristoufek2023,volkovich2023}. These rules can be made explicit, monitored on–chain, and governed transparently.

For traders, the comparison of IV with the MS–GARCH volatility $\sigma_t$ serves as a standard level diagnostic rather than an arbitrage proof. When $\text{IV}<\sigma_t$, a delta–hedged long isolates volatility exposure and, when $\text{IV}>\sigma_t$, the natural short–vol leg must be implemented off–protocol or approximated via liquidity provision, since end–users cannot write on Hegic \citep{Hegic2020,Rahman2022,AndolfattoNaikSchoenleber2024}. Realized outcomes depend on fees, slippage, liquidity depth, and jump risk during rebalancing \citep{akyildirim2020,sebastiano2020,volkovich2023}. For liquidity providers, sustained IV premia (IV$>\sigma_t$) with moderate realized variance raise expected premia retention and sustained discounts (IV$<\sigma_t$) do the opposite. Because the pool is the sole writer, buying options from the pool mainly internalizes part of one’s own short–gamma risk and is not a free gain. Practical risk management as monitoring IV–$\sigma_t$ spreads, utilization, oracle latency, and depth, helps to adjust utilization caps and inventory tolerances when spreads increases and liquidity depth decreases.

Policymakers and standard setters should adopt a disclosure standard for DeFi option AMMs that mandates oracle transparency to publicly report update thresholds, expected latencies, and fallback logic. There should be a requirement of routine reporting of liquidity and utilization, queuing dynamics, and any withdrawal gates. Additionally, the disclosure standard should provide user-facing risk notices on short-maturity convexity and the transaction costs of delta-hedged strategies. Event-triggered updates, e.g., during volatility spikes or oracle incidents, should be required to keep these disclosures relevant. This technology standard can improve price formation, reduce timing and funding uncertainty, and lower the risk that observed price differences are misread as informational inefficiency when they primarily reflect market structure.

It is important to clarify that our price estimates serve as a model-based benchmark rather than a definitive true price. Our valuation method uses the BS model with a regime-sensitive volatility, a practical approach for short-maturity options in emerging environments \citep{venter2020}. However, this model has several limitations. It intentionally omits complex features like jump--diffusion and stochastic volatility over the live time of options. Consequently, in market states where these factors are significant, our calculated option values may be systematically biased.
The identification of this study is cross-sectional and conditional. Therefore, the reported results should be interpreted as local associations. Furthermore, the model abstracts from critical market design features and frictions. We do not account for AMM inventory and fee mechanics, execution costs, collateral and stablecoin risks, or cross-chain bridge delays. As a result, what may appear as mispricings could reflect these microstructure frictions rather than pure risk premia. 
The external validity of our findings is also limited. Significant differences in fees, margin requirements, and settlement conventions between centralized and decentralized exchanges make cross-market comparisons difficult without controlling for microstructure effects. 
The limitations of this study suggest several avenues for future research. Subsequent work could extend the analysis to protocols that permit end-user shorting to test the role of inventory constraints. To better separate risk premia from execution costs, future studies could incorporate real-time, cross-exchange hedging frictions. To assess the stability of the results, it could be of advantage to test alternative regime specifications and look-back windows for the volatility estimation \citep{wu2021,liang2022}.
To sum up, while our evidence provides a coherent benchmark under its stated assumptions, it should not be interpreted as establishing causal effects or unique market prices.

For Hegic protocol, economically meaningful and directionally coherent links between mispricing and order size, moneyness, maturity (for wBTC), volatility (for wBTC), and liquidity, point to design and state variable as first–order drivers of on–chain option pricing. The design recommendations above provide immediately actionable calibration targets for developers, the IV–$\sigma_t$ diagnostic and delta–neutral hedge opportunities provide practical decision rules for traders and liquidity providers. Together, these steps can narrow systematic price deviations while retaining the simplicity and composability that make on–chain option AMMs attractive.

\bibliography{sn-bibliography}
\newpage
\begin{appendices}

\section{Appendix}\label{sec:appendix}
An appendix contains supplementary information that is not an essential part of the text itself but which may be helpful in providing a more comprehensive understanding of the research problem or it is information that is too cumbersome to be included in the body of the paper.

\begin{table}[htbp]
  \centering
  \caption{Descriptive statistics of underlyings}
  \label{TAB:summaryunderlying}
  \begin{tabular}{l
                  S[table-format= 2.4, table-number-alignment=right]
                  S[table-format= 2.4, table-number-alignment=right]}
    \toprule
    & {BTC} & {ETH} \\
    \midrule
    Mean                     &  0.1491 &  0.1849 \\
    Std Deviation            &  3.5875 &  4.6091 \\
    Min                      & -47.9934 & -56.9506 \\
    \SI{25}{\%}-Quartile & -1.3578 & -1.8478 \\
    \SI{50}{\%}-Quartile &  0.0778 &  0.1204 \\
    \SI{75}{\%}-Quartile &  1.7233 &  2.3479 \\
    Max                      & 17.7913 & 23.3480 \\
    Skewness                 & -1.1730 & -1.1068 \\
    Kurtosis                 & 21.4857 & 18.1792 \\
    \bottomrule
  \end{tabular}

\end{table}

\begin{table}[htbp]
  \centering
  \caption{Descriptive statistics of options}
  \label{tab:DescriptiveStats}
  \begin{tabular}{
      l
      S[table-format = 3.4]
      S[table-format = 6.4]
      S[table-format = 7.4]
      S[table-format = 2.2]
      S[table-format = 1.4]}
    \toprule
    & {Amount} & {Premium Paid} & {Strike} & {Maturity} & {Moneyness} \\
    \midrule
    \multicolumn{6}{l}{\textbf{Call options — wBTC} ($n$ = \num{723})} \\
    Mean            & 1.3806 & 699.5291  & 29064.7756 & 17 & 0.9495 \\
    Std Deviation   & 4.5786 & 2253.8008 & 13989.1117 & 21 & 0.0577 \\
    Min             & 0.0000 & 0.0186    & 15574.2470 &  7 & 0.7692 \\
    \SI{25}{\%}-Quartile & 0.0350 &  15.8999 & 19952.6608 &  7 & 0.9091 \\
    \SI{50}{\%}-Quartile & 0.1500 &  47.0000 & 24312.8100 &  7 & 1.0000 \\
    \SI{75}{\%}-Quartile & 0.5000 & 366.7770 & 29814.1561 & 15 & 1.0000 \\
    Max             & 53.5000 & 28075.9242 & 78565.2299 & 90 & 1.0000 \\
    \midrule
    \multicolumn{6}{l}{\textbf{Put options — wBTC} ($n$ = \num{592})} \\
    Mean            & 0.6232 &  559.0408 & 28933.0447 & 19 & 1.0426 \\
    Std Deviation   & 1.5282 & 2236.6310 & 13319.5975 & 20 & 0.0696 \\
    Min             & 0.0001 &    0.1085 & 11362.7045 &  7 & 1.0000 \\
    \SI{25}{\%}-Quartile & 0.0250 &  22.7973 & 20540.6068 &  7 & 1.0000 \\
    \SI{50}{\%}-Quartile & 0.1000 &  70.3125 & 25603.1500 & 10 & 1.0000 \\
    \SI{75}{\%}-Quartile & 0.5000 & 268.8343 & 30303.9775 & 26 & 1.1111 \\
    Max             & 15.0000 & 39551.5471 & 72142.1470 & 90 & 1.4286 \\
    \midrule
    \multicolumn{6}{l}{\textbf{Call options — ETH} ($n$ = \num{1664})} \\
    Mean            &  9.3663 &  673.4381 & 1943.7615 & 22 & 0.9651 \\
    Std Deviation   & 22.2319 & 1894.1741 &  617.8966 & 24 & 0.0567 \\
    Min             &  0.0000 &    0.0000 & 1089.4159 &  7 & 0.7692 \\
    \SI{25}{\%}-Quartile & 0.2500 &  11.7653 & 1572.7411 &  7 & 0.9091 \\
    \SI{50}{\%}-Quartile & 1.1395 &  83.7159 & 1790.0698 & 14 & 1.0000 \\
    \SI{75}{\%}-Quartile & 8.0000 & 494.0209 & 2091.9256 & 26 & 1.0000 \\
    Max             & 264.0000 & 24372.6270 & 4457.6131 & 90 & 1.0000 \\
    \midrule
    \multicolumn{6}{l}{\textbf{Put options — ETH} ($n$ = \num{1111})} \\
    Mean            & 12.4002 &  626.3159 & 1636.4635 & 19 & 1.0672 \\
    Std Deviation   & 33.4195 & 1963.5321 &  458.8686 & 20 & 0.0969 \\
    Min             &  0.0001 &    0.0020 &  811.7060 &  7 & 1.0000 \\
    \SI{25}{\%}-Quartile & 1.0000 &  19.9412 & 1398.3643 &  7 & 1.0000 \\
    \SI{50}{\%}-Quartile & 2.0000 &  99.4667 & 1590.4926 & 10 & 1.0000 \\
    \SI{75}{\%}-Quartile &10.0000 & 474.4329 & 1801.4949 & 21 & 1.1111 \\
    Max             &516.0000 & 34994.2598 & 4043.2200 & 90 & 1.4286 \\
    \bottomrule
  \end{tabular}

\end{table}

\begin{table}[htbp]
 \centering
 \caption{Descriptive statistics of mispricing}
\label{TAB:statsmispricing}
 \begin{tabular}{lcccc}
   \toprule
  & \multicolumn{2}{c}{wBTC} & \multicolumn{2}{c}{ETH} \\
   \cmidrule(lr){2-3}\cmidrule(lr){4-5}
  & {Call options} & {Put options} & {Call options} & {Put options} \\
   \midrule
   Mean                     &  0.17505 &  0.02490 &  0.09730 & -0.06450 \\
   Std Deviation            &  0.67728 &  0.39620 &  0.42850 &  0.35810 \\
   Min                      & -0.99617 & -0.99410 & -0.98750 & -0.99990 \\
 \SI{25}{\%}-Quartile & -0.17070 & -0.20700 & -0.12600 & -0.27320 \\
  \SI{50}{\%}-Quartile &  0.05200 &  0.01050 &  0.06310 & -0.02430 \\
  \SI{75}{\%}-Quartile &  0.31120 &  0.27020 &  0.24370 &  0.13890 \\
  Max                      &  4.58930 &  1.41200 &  4.29980 &  2.05460 \\
  Skewness                 &  2.77620 &  0.33981 &  2.23161 &  0.14577 \\
  Kurtosis                 & 13.82080 &  4.00729 & 14.96550 &  5.24340 \\
  \bottomrule
 \end{tabular}
\end{table}

\begin{table}[htbp]
  \centering
  \caption{Variance inflation factor}
  \label{TAB:vif}
  \begin{tabular}{lcc}
    \toprule
    & \multicolumn{1}{c}{wBTC} & \multicolumn{1}{c}{ETH} \\
    \midrule
    Intercept   & 4.6253 & 4.4292 \\
    Amount      & 1.1222 & 1.0892 \\
    Strike      & 1.4042 & 1.2357 \\
    Maturity    & 1.1429 & 1.1200 \\
    Return      & 1.0434 & 1.0505 \\
    Volume      & 1.2593 & 1.4251 \\
    Volatility  & 1.1175 & 1.3186 \\
    Kind        & 1.0420 & 1.1032 \\
    Type        & 1.2932 & 1.1450 \\
    \bottomrule
  \end{tabular}

\end{table}

\begin{table}[htbp]
  \centering
  \caption{Descriptive statistics of volatility differences}
  \label{TAB:statsvoldifff}
  \begin{tabular}{lcccc}
    \toprule
    & \multicolumn{2}{c}{BTC} & \multicolumn{2}{c}{ETH} \\
    \cmidrule(lr){2-3}\cmidrule(lr){4-5}
    & {Call options} & {Put options} & {Call options} & {Put options} \\
    \midrule
    Mean                      & -0.0194 & -0.0055 & -0.0074 &  0.0484 \\
    Std Deviation             &  0.1432 &  0.1542 &  0.1643 &  0.1849 \\
    Min                       & -0.5074 & -0.4451 & -0.6082 & -0.5378 \\
    \SI{25}{\%}-Quartile & -0.1036 & -0.1145 & -0.0920 & -0.0598 \\
    \SI{50}{\%}-Quartile & -0.0171 & -0.0047 & -0.0276 &  0.0084 \\
    \SI{75}{\%}-Quartile &  0.0627 &  0.0936 &  0.0604 &  0.1269 \\
    Max                       &  0.4749 &  0.7154 &  0.8252 &  0.9127 \\
    Skewness                  &  0.1700 &  0.3819 &  0.4054 &  1.2628 \\
    Kurtosis                  &  3.7762 &  4.4548 &  5.8807 &  6.4725 \\
    \bottomrule
  \end{tabular}
\end{table}

\begin{figure}[htbp]
    \centering
    \caption[PACF]{PACF}
    \begin{subfigure}{.5\textwidth}
        \centering
        \captionsetup{justification=centering}
        \includegraphics[width=\textwidth]{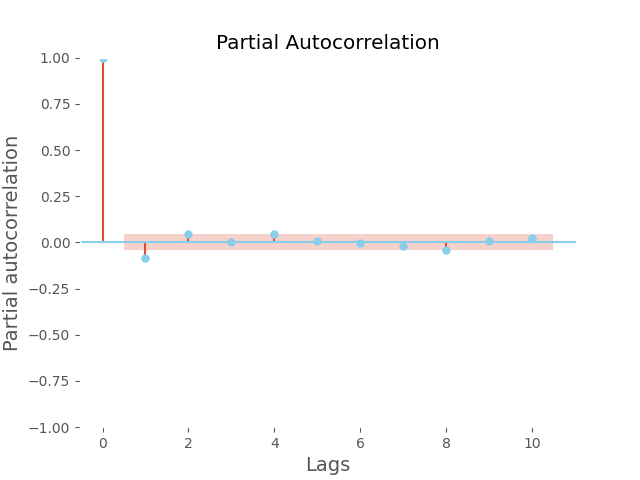}
        \caption{BTC}
        \label{FIG:pacf_btc}
    \end{subfigure}%
    \begin{subfigure}{.5\textwidth}
        \centering
        \captionsetup{justification=centering}
        \includegraphics[width=\textwidth]{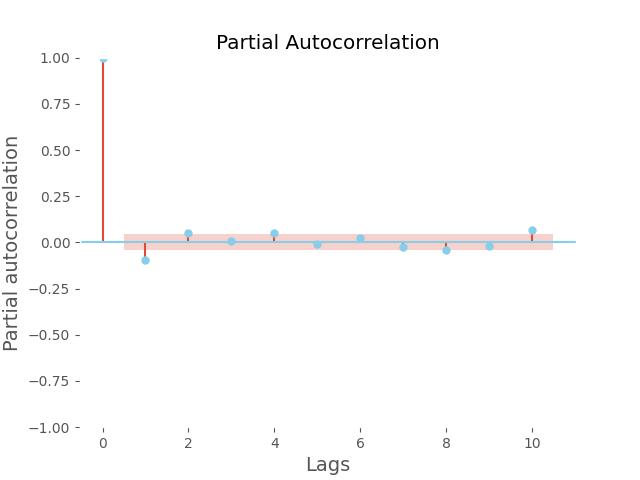}
        \caption{ETH}
        \label{FIG:pacf_eth}
    \end{subfigure}
\end{figure}

\begin{figure}[htbp]
	\centering
	\caption[Correlation]{Correlation matrices.}
	\begin{subfigure}{.5\textwidth}
		\centering
		\includegraphics[width=\textwidth]{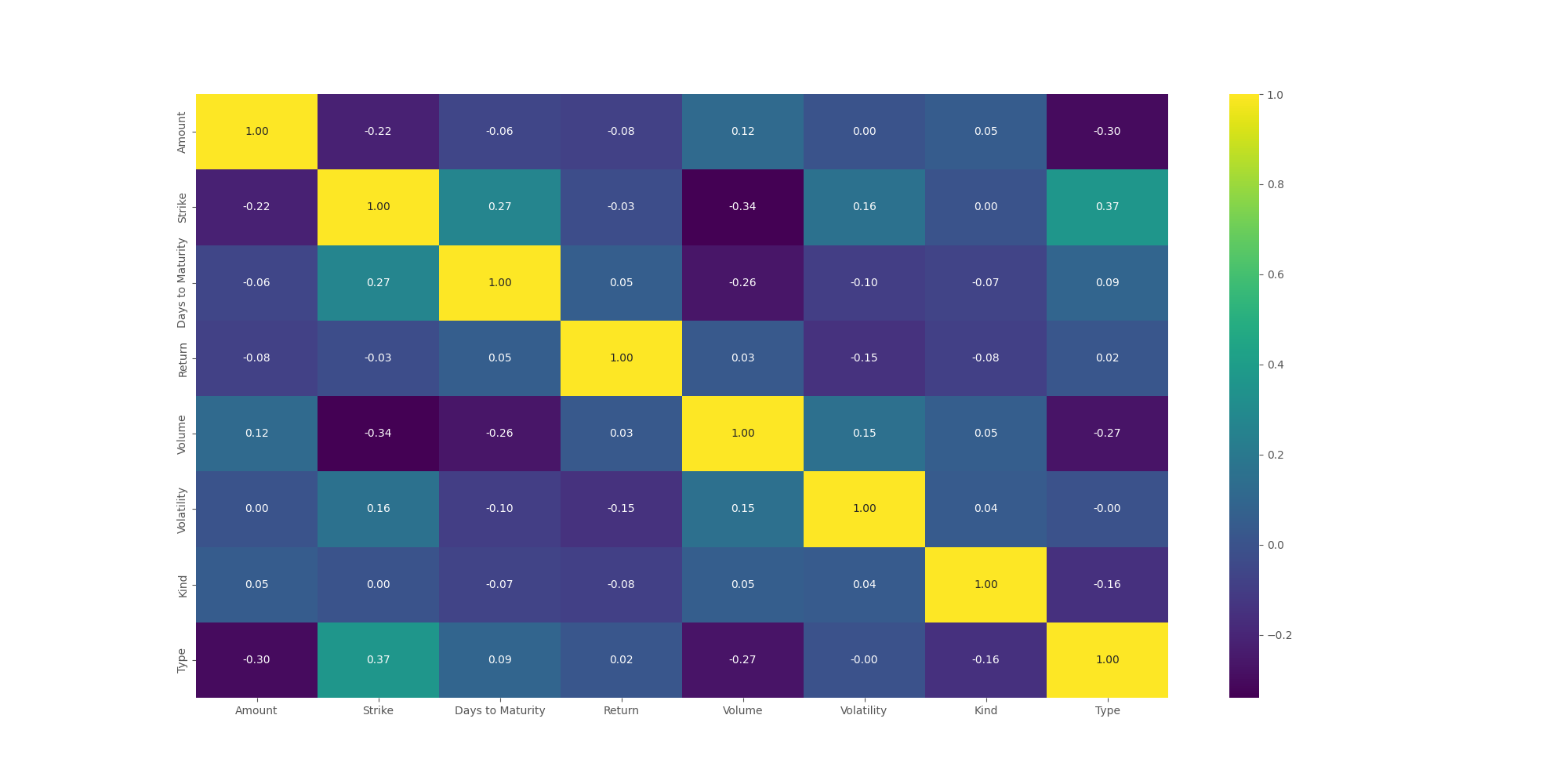}
		\captionsetup{justification=centering}
		\caption{wBTC}
	\end{subfigure}%
	\begin{subfigure}{.5\textwidth}
		\centering
		\includegraphics[width=\textwidth]{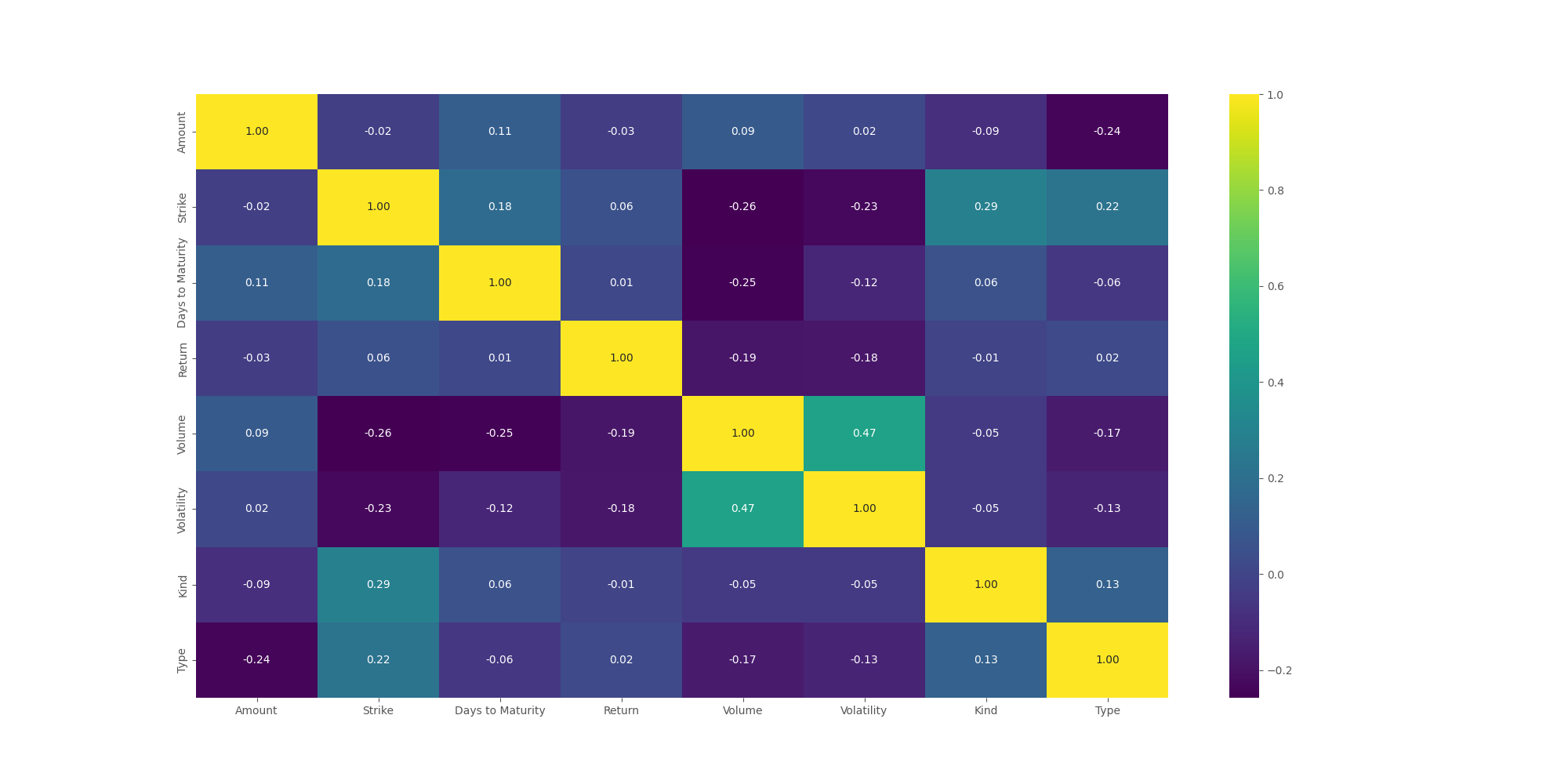}
		\captionsetup{justification=centering}
		\caption{ETH}
	\end{subfigure}
	\label{FIG:correl}
\end{figure}

\end{appendices}


\end{document}